\documentclass[11pt,letterpaper]{article}
\usepackage[margin=1in]{geometry}
\usepackage[T1]{fontenc}
\usepackage{lmodern}
\usepackage{microtype}
\usepackage{amsmath,amssymb,amsthm,mathtools}
\usepackage{booktabs}
\usepackage{enumitem}
\usepackage{xcolor}
\usepackage[ruled,vlined,linesnumbered]{algorithm2e}
\usepackage{tikz}
\usepackage[colorlinks=true,linkcolor=blue!50!black,citecolor=green!35!black,urlcolor=blue!50!black]{hyperref}
\usepackage[capitalize,nameinlink]{cleveref}

\setlist{itemsep=2pt,topsep=4pt}

\newtheorem{theorem}{Theorem}
\newtheorem{lemma}{Lemma}[section]
\newtheorem{proposition}[lemma]{Proposition}
\newtheorem{corollary}[lemma]{Corollary}
\theoremstyle{definition}
\newtheorem{definition}[lemma]{Definition}

\theoremstyle{remark}

\crefname{lemma}{Lemma}{Lemmas}
\crefname{proposition}{Proposition}{Propositions}
\crefname{corollary}{Corollary}{Corollaries}
\crefname{remark}{Remark}{Remarks}
\crefname{example}{Example}{Examples}
\crefname{definition}{Definition}{Definitions}

\newcommand{\F}{\mathbb{F}}
\newcommand{\Fq}{\mathbb{F}_q}
\newcommand{\E}{\mathbb{E}}
\renewcommand{\Pr}{\mathbb{P}}
\newcommand{\Ot}{\widetilde{O}}
\newcommand{\Good}{\mathsf{Good}}
\newcommand{\Bud}{\mathsf{Bud}}
\newcommand{\hist}{\mathfrak{h}}
\newcommand{\polylog}{\operatorname{polylog}}
\newcommand{\dealer}{\operatorname{dealer}}
\newcommand{\Dec}{\operatorname{Dec}}
\newcommand{\Tr}{\operatorname{Tr}}
\newcommand{\ket}[1]{\lvert #1\rangle}
\newcommand{\bra}[1]{\langle #1\rvert}
\newcommand{\ketbra}[2]{\lvert #1\rangle\langle #2\rvert}
\newcommand{\one}{\mathbf{1}}
\newcommand{\lab}[1]{\textsf{#1}}

\crefname{algocf}{Algorithm}{Algorithms}
\Crefname{algocf}{Algorithm}{Algorithms}
\crefname{AlgoLine}{line}{lines}

\newcommand{\QBA}{\textup{\textmd{\textsc{QBA}}}}
\newcommand{\QBAP}{\textup{\textmd{\textsc{QBA:\allowbreak PackedShares}}}}
\newcommand{\CoinPhases}{\textup{\textmd{\textsc{ByzantineAgreementWithCommonCoin}}}}
\newcommand{\LBE}{\textup{\textmd{\textsc{LightestBinElection}}}}

\newcommand{\FinalAg}{\textup{\textmd{\textsc{FinalAgreement}}}}
\newcommand{\DispTr}{\textup{\textmd{\textsc{DisposableTransfer}}}}

\title{Fast Quantum Byzantine Agreement \\ with Nearly Linear Quantum Communication}
\author{Jan Olkowski}
\date{}

\begin{document}
\pagenumbering{roman}
\maketitle
\thispagestyle{empty}
\begin{abstract}
\noindent
We consider synchronous binary Byzantine agreement among $n$ processes connected by authenticated point-to-point classical and quantum channels. The adversary is computationally unbounded, sees all classical information, adaptively corrupts up to $t\le(1/3-\varepsilon)n$ processes, and may suppress messages sent by a process in the round in which it is corrupted.

We give a quantum algorithm that solves Byzantine agreement in this model with probability $1$. It uses $O(\log^4 n)$ rounds, $O(n\log^8 n)$ qubits, and $O(n^2\log^8 n)$ classical bits in expectation, with constants depending on $\varepsilon$. This reduces the quantum communication of the protocol of Ben-Or and Hassidim (STOC~2005), which achieves $O(1)$ expected rounds but whose Byzantine coin implementation requires $\Omega(n^4\log n)$ qubits and classical bits.

We also extend the quadratic message lower bound of Abraham et al.\ (PODC~2019) to quantum channels, obtaining an $\Omega(t^2)$ lower bound on worst-case expected total communication. For $t=\Theta(n)$, the communication and round complexity of our algorithm are therefore optimal up to polylogarithmic factors.

Our algorithmic construction builds on the election tree of King and Saia (JACM~2011), replacing its hidden random words with quantum shares. The main difficulty is transferring these words under adaptive corruption. We develop and analyze a scheme in which each holder coherently re-encodes its shares and sends them up the tree through sparse communication graphs, leaving no old input for later corruptions to reveal. We prove that repeated transfers preserve privacy and classical reconstruction despite adaptive Byzantine failures and sparse communication. Finally, we propose a quantum packing techniques to combine all shares of one word held by a process into a single block. Applying this technique on each level on the election tree yields the nearly linear quantum communication bound.
\end{abstract}
\newpage
\pagenumbering{arabic}
\section{Introduction}\label{sec:intro}

In Byzantine agreement, processes must decide a common value even when some behave arbitrarily~\cite{PeaseShostakLamport80}. Randomized protocols can reach agreement quickly if they have enough \emph{common coins}: random values that honest participants obtain identically and that the adversary cannot predict before they are used~\cite{BenOr83,Rabin83}. Producing these coins is difficult against a full-information adaptive adversary. Such an adversary observes classical random choices as they are made and can use them to choose subsequent corruptions. In the synchronous setting considered here, classical protocols require $\Omega(t^2/(n\log n))$ expected rounds~\cite{Efron26}.

Quantum communication makes it possible to keep a coin in a distributed superposition until it is needed. The adversary still sees all classical information, but cannot inspect the registers of uncorrupted processes. Ben-Or and Hassidim~\cite{BenOrHassidim05} used this property to obtain agreement in $O(1)$ expected rounds for $t<n/3$. Li, Sun and Zhu~\cite{LiSunZhu24} later developed a compiler from classical private-channel protocols to quantum full-information protocols.

Our focus is the communication required for fast agreement. King and Saia~\cite{KingSaia11} use $\Ot(n^{3/2})$ bits, but their model has private channels and does not allow the adversary to suppress messages already in transit. Suppression by corrupting a sender makes quadratic communication necessary for randomized classical agreement~\cite{AbrahamCDNPRS19}. On the quantum side, the specified Byzantine coin of Ben-Or and Hassidim uses $\Omega(n^4\log n)$ qubits per flip, as counted in \cref{app:bh}. Nearly linear quantum communication has been obtained for crash failures~\cite{HajiaghayiKO23,HajiaghayiKO24}. We ask:
\begin{quote}
\emph{How much communication does Byzantine agreement against a full-information adaptive adversary need, and how much of it has to be quantum?}
\end{quote}
We show that quadratic total communication is necessary and, up to polylogarithmic factors, sufficient, and that all but a nearly linear part of it can be classical. Our algorithm tolerates $t\le(1/3-\varepsilon)n$ Byzantine processes, for any constant $0<\varepsilon<1/3$, and sends $O(n\log^8n)$ qubits and $O(n^2\log^8n)$ classical bits in $O(\log^4n)$ rounds, all in expectation (\cref{thm:packed}). The matching lower bound extends that of Abraham et al.~\cite{AbrahamCDNPRS19} to quantum channels (\cref{thm:lower,cor:optimal}).

\subsection{Model and complexity measures}\label{sec:model}
The system consists of $n$ processes with known identifiers $P=[n]$, each with an input bit, classical memory, and quantum registers. Every pair has reliable authenticated point-to-point classical and quantum channels. We assume no signatures, trusted setup, broadcast primitive, or initially shared entanglement. Processes operate in synchronous rounds, and messages arrive at the end of the sending round. Gates and channels are ideal; local computation, storage, and per-round bandwidth are unrestricted. Public committees and forwarding graphs are chosen by exhaustive search, so we claim no polynomial bound on local computation.

The computationally unbounded adversary permanently corrupts at most $t\le(1/3-\varepsilon)n$ processes, where $0<\varepsilon<1/3$ is fixed. It observes all classical states, messages, and measurement outcomes, and chooses corruptions adaptively, including within a round. It is rushing: it sees honest messages before choosing its own. It may apply arbitrary joint quantum instruments to corrupted registers and its memory, but cannot access uncorrupted registers. Corruption choices are classical. By corrupting a sender during a round, it may seize, alter, or suppress that sender's outgoing messages; this is its only way to interfere with messages in transit. Channel-use events and their endpoints are visible before delivery. A sender is treated as honest for delivery only if it remains honest through the round, and a quantum transfer is complete only at the end of its round.

A process is \emph{honest at a boundary} if it has not yet been corrupted, and \emph{correct} if it is never corrupted. A boundary may occur inside a round. Agreement requires all correct processes to decide the same bit; validity requires decision $v$ if all correct inputs are $v$. Both properties hold on every execution. Every correct process must decide and halt with probability~1.

We count qubits $Q$ and classical bits $C$ sent while honest, and rounds $R$, through the last correct halt. Communication includes setup and replies to faulty requesters. Sending to several recipients incurs one transmission per recipient. Every channel use costs at least one bit or qubit, including an empty message, so channel-use events convey no information for free. Our upper bounds also hold when all prescribed transmissions of a round are charged even if the sender is corrupted later in that round. Expectations are uniform over inputs and adversaries.

We write $L=\lceil\log_2n\rceil$. The notation $\Ot$ hides polylogarithmic factors in $n$; constants in $O_\varepsilon$ and $O_{\varepsilon,c}$ may depend on the indicated parameters.

\subsection{Main results}\label{sec:ov-results}

Our main result is an agreement algorithm, \QBAP{}, with nearly linear quantum communication and polylogarithmic expected running time.

\begin{theorem}[\cref{sec:proof-packed}]\label{thm:packed}
Fix $0<\varepsilon<1/3$. For every $n$ and every $t\le(1/3-\varepsilon)n$, algorithm \QBAP{} satisfies agreement and validity on every execution and terminates with probability~1. Its expected costs are
\[
\E[Q]=O_{\varepsilon}(n\log^8n),\qquad \E[C]=O_{\varepsilon}(n^2\log^8n),\qquad \E[R]=O_{\varepsilon}(\log^{4}n).
\]
\end{theorem}

We track the exponents of $\log n$ but have not optimized the constants.

Our second result extends the message-suppression argument of Abraham et al.~\cite{AbrahamCDNPRS19} to quantum channels. It uses the adversary's ability to corrupt a sender during a round and suppress its outgoing messages.

\begin{theorem}[\cref{sec:lower}]\label{thm:lower}
Consider binary Byzantine agreement in the model of \cref{sec:model}, except that $t$ may be any integer with $1\le t\le n-2$, and any algorithm that satisfies agreement and validity on every execution and terminates with probability~$1$. Let $K=\sup_{\mathbf x,\mathcal A}\E_{\mathbf x,\mathcal A}[C+Q]$, where the supremum is over input vectors $\mathbf x$ and adversaries $\mathcal A$ that corrupt at most $t$ processes. Then $K\ge\frac14\lfloor(t+1)^2/4\rfloor$.
\end{theorem}

\begin{corollary}[\cref{sec:lower}]\label[corollary]{cor:optimal}
Let $t=\Theta(n)$. Every algorithm as in \cref{thm:lower} has worst-case expected total communication $\sup_{\mathbf x,\mathcal A}\E_{\mathbf x,\mathcal A}[C+Q]=\Omega(n^2)$. If moreover its worst-case expected quantum communication $\sup_{\mathbf x,\mathcal A}\E_{\mathbf x,\mathcal A}[Q]$ is $o(n^2)$, then its worst-case expected classical communication $\sup_{\mathbf x,\mathcal A}\E_{\mathbf x,\mathcal A}[C]$ is $\Omega(n^2)$. Hence the total and the classical communication of \QBAP{} are optimal up to polylogarithmic factors.
\end{corollary}

\begin{table}[t]
\centering
\footnotesize
\setlength{\tabcolsep}{4pt}
\renewcommand{\arraystretch}{1.15}
\begin{tabular}{@{}lllllll@{}}
\toprule
Work & Channels / view & Faults & $t$ & Exp.\ rounds & Qubits & Classical\\
\midrule
\cite{KingSaia11}${}^{a}$ & classical, private & Byzantine & $(\frac13-\varepsilon)n$ & $\polylog n$ & -- & $\Ot(n^{3/2})$ bits\\
\cite{DufoulonPandurangan25}${}^{b}$ & classical, full info & Byzantine & $<n/3$ & $O(\min\{\frac{t^2\log n}{n},\frac{t}{\log n}\})$ & -- & $\Ot(nt^2)$ messages\\
\cite{BenOrHassidim05} & quantum, full info & Byzantine & $<n/3$ & $O(1)$ & $\Omega(n^4\log n)$ & $\Omega(n^4\log n)$\\
\cite{HajiaghayiKO24} & quantum, full info & crash & $<n/3$ & $O(1)$ & $O(n^{1+\delta})$ & $O(n^{1+\delta})$ bits\\
\midrule
\Cref{thm:packed} & quantum, full info & Byzantine & $(\frac13-\varepsilon)n$ & $O(\log^{4}n)$ & $O(n\log^8n)$ & $O(n^2\log^8n)$ bits\\
\Cref{cor:optimal} & quantum, full info & Byzantine & $\Theta(n)$ & any & \multicolumn{2}{l}{$\E[C+Q]=\Omega(n^2)$}\\
\bottomrule
\end{tabular}
\caption{Synchronous agreement against an adaptive adversary. ${}^{a}$Agreement only with high probability, against an adversary that cannot suppress messages in transit. ${}^{b}$Message complexity $O(\min\{nt^2\log n,\,n^2t/\log n\})$; agreement only with high probability.}
\label{tab:results}
\end{table}

\Cref{tab:results} compares the communication and time bounds. To our knowledge, no previous Byzantine agreement protocol against a full-information adaptive adversary has total communication $\Ot(n^2)$. The quantum protocol of Ben-Or and Hassidim achieves constant expected time at a higher communication cost. Our protocol instead uses $O(\log^4 n)$ expected rounds. For $t=\Theta(n)$, even this polylogarithmic time is unavailable to classical protocols in the same model, by the lower bound of Efron~\cite{Efron26}.

\subsection{Relation to earlier constructions}\label{sec:contrib}

We adapt the election tree and fixed committees of King and Saia~\cite{KingSaia11}, together with Feige's lightest-bin rule~\cite{Feige99}, to randomness hidden in quantum superpositions, following Ben-Or and Hassidim~\cite{BenOrHassidim05}. The main difficulty is preserving this randomness as its holders change. King and Saia erase old shares; we introduce disposable quantum shares, whose coherent transfer consumes the old inputs. Our main technical contribution is proving that repeated transfers preserve privacy and permit consistent classical reconstruction under adaptive Byzantine corruption.

Replacing erasure is essential: the reduction of Li, Sun and Zhu~\cite{LiSunZhu24} excludes protocols that use it and therefore does not apply to King and Saia's construction. But protecting the transferred randomness is only part of the problem. Unlike King and Saia's adversary, ours can corrupt a sender during a round and suppress its outgoing messages without reading their quantum contents. Under our agreement guarantees, this ability forces $\Omega(n^2)$ worst-case expected total communication when $t=\Theta(n)$, even with quantum channels (Theorem~\ref{thm:lower}). This strictly differentiate our model from King and Saia's in which $\Tilde O(n^{3/2})$ communication complexity is achieved. 

To obtain the quantum bound, we pack all registers of a word held by one process into a single transfer block, preventing their number from multiplying across election levels. Polynomial packing is an established technique~\cite{SenthoorSarvepalli22}; our contribution is its adaptation to disposable shares and the proof that privacy and reconstruction survive these packed transfers. Together with geometrically growing committees, this yields $O_\varepsilon(n\log^8 n)$ expected quantum communication.

\subsection{Related work}\label{sec:related}

\paragraph{Randomized agreement and common coins.}
Since the work of Rabin~\cite{Rabin83}, randomized Byzantine agreement has been reduced to the generation of common coins. Ben-Or~\cite{BenOr83} uses local coins, which are common only with exponentially small probability. Feldman and Micali~\cite{FeldmanMicali97} generate, in a constant number of rounds, a coin that is common with constant probability, against an adaptive adversary that cannot read private channels.

\paragraph{Round complexity against a full-information adaptive adversary.}
Classical synchronous protocols against this adversary need $\Omega(t/\sqrt{n\log n})$ expected rounds~\cite{BarJosephBenOr98} in case of weaker \textit{crash} failures, and $\Omega(t^2/(n\log n))$ for Byzantine failures, by a recent paper of Efron~\cite{Efron26}. For $t=\Theta(n)$, Hajiaghayi, Kowalski and Olkowski~\cite{HajiaghayiKO22} matched the former lower bound up to polylogarithmic factors. The protocol of Dufoulon and Pandurangan~\cite{DufoulonPandurangan25} matches the latter up to polylogarithmic factors, using $\Ot(nt^2)$ messages. For asynchronous full-information agreement with $t=\Theta(n)$, Li, Sun and Zhu~\cite{LiSunZhu24} cite an $\Omega(n)$-round lower bound based on Attiya and Censor~\cite{AttiyaCensor08}. Huang, Pettie and Zhu~\cite{HuangPettieZhu24} gave polynomial-time asynchronous full-information protocols with near-optimal and then optimal resilience.

\paragraph{Communication complexity of Byzantine agreement.}
Dolev and Reischuk~\cite{DolevReischuk85} showed that deterministic agreement needs $\Omega(t^2)$ messages. Abraham et al.~\cite{AbrahamCDNPRS19} extended the quadratic bound on the number of messages to randomized protocols against a strongly adaptive adversary, and \cref{thm:lower} extends their argument to quantum channels. Against an adaptive adversary that cannot remove messages in transit, King and Saia~\cite{KingSaia11} send $\Ot(n^{3/2})$ bits, using Feige's bin selection~\cite{Feige99} and erasure of old shares. The compiler of Li, Sun and Zhu excludes protocols that erase, so it does not apply to theirs~\cite{LiSunZhu24}. With cryptographic setup, subquadratic communication is possible even asynchronously; see, for example, Blum, Katz, Liu-Zhang and Loss~\cite{BlumKLL20}. We use no cryptographic assumptions.

\paragraph{Quantum agreement protocols.}
Fitzi, Gisin and Maurer~\cite{FitziGisinMaurer01} used entanglement to solve a three-party detectable variant of broadcast. Ben-Or and Hassidim~\cite{BenOrHassidim05} replaced the classical secrets of Feldman and Micali's coin, one for each ordered pair of processes, with quantum superpositions shared by quantum verifiable secret sharing~\cite{CrepeauGottesmanSmith02}; \cref{app:bh} counts the qubits of one coin flip. Li, Sun and Zhu~\cite{LiSunZhu24} generalized this idea into a compiler from classical private-channel protocols to quantum full-information protocols, which preserves resilience, rounds and the expected number of messages, but in general not the number of communication bits, and used it to obtain constant-round asynchronous protocols.

Hajiaghayi, Kowalski and Olkowski~\cite{HajiaghayiKO24} reduced the quantum and classical communication of crash-tolerant quantum consensus to $O(n^{\delta})$ per process in $O(1)$ expected rounds, or to polylogarithmic per process in polylogarithmic rounds, both amortized and with high probability. Their technique combines a quantum weak global coin with sparse random communication patterns. Earlier work by Chlebus, Kowalski and Strojnowski~\cite{ChlebusKS10} had already explored sparse quantum communication for crash-tolerant consensus.

\paragraph{Secret sharing and other ingredients.}
Shamir's scheme~\cite{Shamir79} and its Reed--Solomon view~\cite{ReedSolomon60,McElieceSarwate81} underlie our words. Threshold quantum secret sharing was introduced by Cleve, Gottesman and Lo~\cite{CleveGottesmanLo99} and developed by Gottesman~\cite{Gottesman00}. Our shared values are classical uniform values held in superposition. Therefore we need only privacy and classical decoding, and not recovery of an arbitrary quantum state. Packing several secrets into one polynomial goes back to Franklin and Yung~\cite{FranklinYung92}, and its quantum ramp analogue is studied by Senthoor and Sarvepalli~\cite{SenthoorSarvepalli22}; we use packing in \QBAP{}. We use Hoeffding and Chernoff bounds in the probabilistic existence arguments~\cite{Hoeffding63,MitzenmacherUpfal17}.
\paragraph{Organization.}
\Cref{sec:results} explains the construction. The formal argument has three parts: the protocol specification and randomness requirements in \cref{sec:interface}, agreement assuming that interface in \cref{sec:classical}, and its quantum implementation in \cref{sec:moving}. \Cref{sec:proof-packed} chooses parameters and proves the complexity bounds. The lower bound is independent of the construction and appears in \cref{sec:lower}.

\paragraph{AI methodology statement}
The construction and analysis underlying Theorem~\ref{thm:packed} were developed through multiple rounds of interaction between the author and ChatGPT 6 Astra Medium. The author initially prompted the model to combine the approach of King and Saia~\cite{KingSaia11} with quantum secret sharing. After the model proposed an initial construction with quadratic quantum communication, the author directed it toward a tree of elections with gradually increasing committee sizes, leading to a construction with lower quantum communication.

The model subsequently proposed and developed an application of established packing techniques~\cite{SenthoorSarvepalli22} to reduce quantum communication further. This application of packing was not known to the author before these interactions and was developed by the model without the author supplying that idea. The resulting construction achieves the $O_\varepsilon(n\log^8 n)$ expected quantum communication bound stated in Theorem~\ref{thm:packed}.

The model was also used extensively to develop and revise the exposition of the results. The author takes full responsibility for the accuracy, correctness, and attribution of all content in the paper.

\section{Overview of the construction}\label{sec:results}

Our construction separates the use of randomness from its implementation. We first explain how fresh common coins lead to agreement and how a tree of elections selects a short list of sources for these coins. We then explain how quantum shares keep the unused randomness hidden as it moves through the tree. The main difficulty is to preserve this protection under adaptive corruption without allowing the number of shares to grow at every transfer.

\subsection{Reducing agreement to fresh common coins}\label{sec:ov-coins}

The agreement procedure is classical. Its participants maintain estimates and exchange votes in successive phases. A \emph{fresh common coin} is a bit that all honest participants obtain identically and that is uniform given the adversary's information when the votes of the phase have been fixed. Each such coin gives a probability of at least $1/2$ that all honest estimates become equal. Once they are equal, subsequent phases preserve them, even if later coins are biased or inconsistent. Thus the task is to provide enough fresh common coins at low communication cost.

Consider a committee of $k$ processes with at most $a=\lfloor(k-1)/3\rfloor$ corrupted members. In each phase, every member sends its current estimate to the others. A member proposes $v$ if at least $k-a$ received estimates equal $v$, and proposes $\bot$ otherwise. Two sets of $k-a$ senders intersect in at least $a+1$ members, so honest members cannot propose opposite values. A member that receives at least $a+1$ proposals of $v$ records $v$ as forced. After these proposals have arrived, at the \emph{proposal deadline}, the participants open the coin. Each member adopts its forced value if it has one, and the coin otherwise.

All forced values at honest members are the same. A fresh common coin equals this value with probability $1/2$, aligning all honest estimates; if no honest member is forced, either coin outcome aligns them. The coin must remain hidden until the proposal deadline. Otherwise the adversary could use its value when choosing faulty votes, defeating this argument. Each phase takes four rounds, including the request and reconstruction of the coin.

\paragraph{Lemma~\ref{lem:enough} (informal).}
While at most $a$ members are corrupted, the procedure preserves every bit on which all honest members agree. If $g$ phases provide fresh common coins, the probability that the committee remains within this corruption bound but its honest members disagree on some of $u$ output bits is at most $u2^{-g}$.

One coin bit per phase suffices for all $u$ components: apply the alignment argument to each component and take a union bound. We use this procedure, \CoinPhases{}, both within committees to select sources of randomness and among all processes to agree on an input value. The quantum part of the construction supplies the coins; it does not replace these classical votes.

\subsection{Selecting randomness through committee elections}\label{sec:ov-tree}

The simplest way to supply fresh common coins would be to let each process distribute one hidden random word and open these words in a fixed order. Suppose for now that an honestly supplied word remains hidden until opening and reconstructs identically at honest processes. Among $t+1$ distinct suppliers, at least one is honest, giving a constant probability of aligning the estimates. The adversary can, however, corrupt the first $t$ suppliers in the public order and delay the first useful coin by a linear number of phases. We therefore seek a short list containing enough honest sources of randomness, without identifying them individually.

Following King and Saia~\cite{KingSaia11}, the processes select this list through a tree of committees. Initially, they distribute their candidate randomness to small committees near the leaves. Each committee elects a short list and passes the winners to a larger parent committee. The parent combines the lists from its children and repeats the election. Committee sizes grow toward the root, whose committee consists of all processes. The root uses the surviving randomness for agreement on the input values. The tree has logarithmic height, and each election forwards only a polylogarithmic number of candidates, while losing little of the proportion supplied honestly. Formally, its leaves are initial candidate lists and each internal node represents a committee; committees can overlap.

An election itself needs coins, so a winning candidate must provide randomness for several stages. The processes therefore prepare each candidate as an \emph{array} of independent words, with separate words for successive election levels and a final word for input agreement. Each array has a public identifier. At a given level, a committee opens only the words assigned to its election; the winners' remaining words stay unopened and are transferred to the parent. An array is \emph{good} if its dealer remains uncorrupted through its initial distribution. The dealer assignments ensure an initial good fraction that exceeds one-half by a fixed positive amount. Later corruption of a dealer does not change this classification: the sharing scheme must protect the words after they have left it.

Consider an election whose committee receives $b$ child lists, each containing $w$ positions. It must select $w$ of these $r=bw$ candidates. Using Feige's lightest-bin rule~\cite{Feige99}, it opens a level-specific \emph{bin word} of each array to assign the array to one of $b$ bins. The members then run \CoinPhases{} to agree on the identifier/bin pairs, using the separate coin word of the candidate in position $j$ for phase $j$. Once they agree, they retain every entry of a lightest bin and fill their output to $w$ entries in a fixed public order. Words used for bin assignment, agreement, and later levels are distinct.

For good arrays whose sharing guarantees hold, the bin assignments are independent and uniform. With high probability, every bin receives close to a $1/b$ fraction of these arrays. Faulty candidates may choose their bins after seeing the good assignments, but they cannot remove a good array from a bin. A lightest bin contains at most $r/b=w$ entries, so all its good arrays survive, and filling the output to $w$ entries preserves nearly the original good fraction. Enough good candidates also provide enough fresh coin phases for the committee to agree on the pairs. The one-election bound in \cref{lem:election} accounts for disagreement and bin imbalance.

Some committees may receive too few good arrays, and an adaptive adversary may concentrate corruptions in others. We choose the committees so that every allowed fault set overloads only a small fraction of each level. The proof then counts good arrays across the whole level, retaining only those forwarded through successful elections: the relevant committees stayed within their corruption budgets through delivery and their honest members agreed on the lists. Overloaded committees and elections with too few such arrays contribute a small loss to this count; the remaining elections lose only a small fraction through bin imbalance. These losses are chosen small enough that their sum over the tree leaves a majority of good arrays at the root. The committee and level-counting arguments are given in \cref{lem:committees,lem:induction}.

\paragraph{Proposition~\ref{prop:root} (informal).}
Assume that the relevant unopened words remain jointly uniform and hidden from the adversary, and that honest requesters reconstruct the same values. With the stated parameter choices, all honest processes obtain the same root list with high probability, and a majority of its arrays are good. Their final words have not been used in the elections and provide fresh common coins for the agreement procedure of \cref{sec:ov-coins}. The formal procedure adds a decision rule and delayed halting to turn alignment into safe termination. If an attempt does not finish, undecided processes continue with fresh arrays; each attempt is an \emph{epoch}. Algorithm~\ref{alg:qba} specifies this repetition. It remains to provide the assumed privacy and reconstruction guarantees while transferring the unopened words between committees.

\paragraph{Pseudocode conventions.}\label{sec:code-conventions}
All algorithms describe local actions; each process may participate in several committees. Sending to a committee means one point-to-point message per other member, with the sender's own vote counted locally. Missing or malformed votes count as $\bot$. Threshold tests choose $0$ if both bits qualify. Identifier $0$ denotes an empty position: it generates no request and supplies word value and bin $0$. Missing readouts are replaced by $0$, and failed decoding returns $0$. The prescribed rounds still elapse when a provider is empty; no messages are sent for that request. Holders enforce the authorization and per-requester quotas of \cref{sec:communication} throughout elections and final agreement.

For $h=0$, the level loop is empty: words are distributed directly to $P$ and the root list is the public initial list. At the root there is no parent transfer. A process that has decided remains active, including its holder duties, until the delayed-halt rule of \FinalAg{} permits it to stop. Its decision state persists across epochs.

\begin{algorithm}[t]
\caption{\QBA{} (process $p$)}\label{alg:qba}
\KwIn{input bit $v_p$}
$e_p\gets v_p$\;
\For{each epoch, until $p$ halts}{
  prepare and distribute fresh words for every array dealt by $p$\nllabel{ln:sel-qba-deal}\;
  \For{$\ell\gets1$ \KwTo $h$}{
    run \LBE{} concurrently in every level-$\ell$ committee containing $p$\nllabel{ln:sel-ts-elect}\;
    \tcp{Throughout these elections, holders serve scheduled requests concurrently.}
    during delivery, transfer the unopened words in each local output list to its parent\nllabel{ln:sel-ts-hold}\;
  }
  $A\gets$ the local root list\;
  $e_p\gets\FinalAg(e_p,A)$\tcp*{may decide and halt}
}
\end{algorithm}

\subsection{Preserving randomness across election levels}\label{sec:ov-storage}

The selection argument assumes that a good array's unused words stay hidden and reconstruct consistently. We now explain how to provide these properties. There are two distinct requirements: a dealer must distribute randomness without revealing its value, and later committees must receive its shares without leaving vulnerable copies behind.

\paragraph{Quantum sharing.}
A full-information adversary sees every classical random choice. Following Ben-Or and Hassidim~\cite{BenOrHassidim05}, we instead distribute a superposition of classical sharings. In the simplest example, a dealer prepares a uniform superposition of degree-at-most-$t$ polynomials over a field $\Fq$, with $q=2^d>n$, and sends one evaluation register to each process. Measuring all registers produces Shamir's sharing of a uniform field element, namely the polynomial's constant coefficient. We call this element the word's value and each evaluation register a share.

Any $t$ shares reveal no information about the value. When the word is needed, its holders measure their registers and send the results to the requesters. Reed--Solomon decoding recovers the value despite up to $t$ wrong or missing reports, since $n>3t$. Opening uses quantum registers, but reconstruction communicates only classical values. The share encoder proves these properties in \cref{lem:encoder}. The single-block case is recorded in \cref{cor:single-block}.

We do not verify a dealer's state. A faulty dealer may supply arbitrary registers, and its word may reconstruct differently at different requesters. The selection analysis already allows such words. For an honest dealer, the preparation leaves no retained copy of the input, so later corruption of that dealer does not reveal the word.

\paragraph{Storage within committees.}
Sharing every word globally would implement the required randomness, but distributing roughly $n$ arrays would already cost roughly quadratic quantum communication. A word is used only by the committees that elect its array. We therefore share it initially with its first committee, transfer it to the parent if the array wins, and repeat. The remaining words follow the elected identifiers upward.

This saves communication because committees are small near the leaves. As their sizes increase, the number of nodes decreases. However, moving shares creates two problems that the global construction avoids. Former holders must retain no old shares that later corruptions could reveal, and repeated transfers must not multiply the number of registers. Disposable transfers address the first problem; packing addresses the second.

\subsection{Quantum sharing and disposable transfers}\label{sec:ov-reduce}

\paragraph{Transfer procedure.}
Suppose a holder must send its share to the next committee. Sending an unprotected value would expose it to a corrupted recipient. Instead, the holder re-encodes it into several new shares and sends them to distinct members of the parent committee. A fixed bipartite forwarding graph specifies the recipients. Its outgoing degree is $D_\ell$ at level $\ell$, and each recipient has at most $J$ incoming edges. Both bounds are $O_\varepsilon(\log n)$.

Classically, re-sharing a value while retaining the old value would not solve the problem: the adversary could corrupt former holders after each transfer and accumulate their old shares. King and Saia use erasure to prevent this. Our holders instead apply the encoding coherently and send all resulting registers. The input registers are consumed by the transformation; no old input remains locally after the transfer round. A self-addressed output is a new encoded share. We call this use of the input shares \emph{disposable}.

\paragraph{Share encoder.}
Write $\ket{+_q}=q^{-1/2}\sum_{z\in\Fq}\ket z$ for a uniform superposition over the field. 
The encoder $V_{D,\sigma,p}$ takes $p$ input registers and $\sigma$ fresh uniform registers. It treats the inputs as the lowest $p$ coefficients of a polynomial and the fresh registers as the next $\sigma$ coefficients, then produces its evaluations at $D$ distinct nonzero points. This is done coherently: neither the inputs nor the fresh registers are measured. A Vandermonde transformation, with zero padding, implements the map, and the holder sends all $D$ outputs.

Two properties of this encoder are used throughout. If $D\ge p+2\sigma$, any $\sigma$ outputs carry no information about the inputs, even when the inputs are entangled with an arbitrary reference system. If $D\ge p+3\sigma$, measuring the outputs and decoding corrects $\sigma$ wrong or missing values and recovers the $p$ input values. The first property protects privacy; the second permits classical reconstruction after several transfers. Their precise statements and proof appear together in \cref{lem:encoder}.

\paragraph{Distribution and transfer conventions.}\label{sec:transfer-conventions}
During initial distribution, the dealer applies $V_{k_1,\sigma(k_1),1}$ to a fresh $\ket{+_q}$ for every word and sends output $i$ to member $i$ of its first committee. For $h=0$, that committee is $P$ and $k_1=n$. During a transfer, missing or malformed input registers are replaced by $\ket0$, and additional zero registers pad the input to $p_\ell$. No word is packed together with another word.

If both the sending and receiving committees are $P$, the holder retains its registers and sends nothing; Algorithm~\ref{alg:transfer} is skipped. Otherwise the holder forwards only the arrays in its own local output list, even if committee members disagree. At the receiver, at most one register is accepted per authorized array, word, and incoming edge; all additional registers are discarded. This cap applies independently of whether the election succeeds.

\begin{algorithm}[t]
\caption{\DispTr{} (one holder $i$, one transfer round)}\label{alg:transfer}
\KwIn{holder $i$'s output list $A$; registers of its words; forwarding edges}
\KwOut{new shares sent to the parent committee}
\ForEach{nonempty array identifier $x$ in $A$\nllabel{ln:dt-own}}{
  \ForEach{unopened word $\omega$ of array $x$}{
    assemble its registers in incoming-edge order, padding to $p_\ell$ inputs\nllabel{ln:dt-pad}\;
    apply $V_{D_\ell,\sigma(D_\ell),p_\ell}$ once to the entire input block\nllabel{ln:dt-packed}\;
    send all outputs, one along each forwarding edge; retain no old input\nllabel{ln:dt-send}\;
  }
}
\end{algorithm}

\begin{figure}[t]
\centering
\begin{tikzpicture}[x=1cm,y=1cm,
  mem/.style={circle,draw,minimum size=4.2mm,inner sep=0pt},
  bad/.style={mem,fill=red!30},
  blk/.style={draw,rounded corners=2pt,fill=blue!7,minimum width=17mm,minimum height=6.5mm,inner sep=2pt}]
\foreach \i in {0,...,10}{\node[mem] (p\i) at (\i*0.85,3.1) {};}
\node[bad] at (p3) {}; \node[bad] at (p8) {};
\node[anchor=west] at (9.5,3.1) {$C_{v'}$, level $\ell+1$};
\foreach \i in {0,...,5}{\node[mem] (c\i) at (2.1+\i*0.85,0) {};}
\node[anchor=west] at (9.5,0) {$C_v$, level $\ell$};
\node[mem,fill=blue!20] (u) at (c2) {};
\node[blk] (V) at (3.8,1.45) {$V_{D,\sigma,p}$};
\draw[->] (u) -- node[left,font=\footnotesize,align=right]{$\le J$ registers,\\ padded to $p$} (V.south);
\draw[->,dashed] (5.3,1.45) node[right,font=\footnotesize]{$\sigma$ fresh $\ket{+_q}$} -- (V.east);
\foreach \j in {1,3,4,6,7}{\draw[->] (V.north) -- (p\j.south);}
\node[font=\footnotesize,anchor=east] at (2.0,2.3) {$D$ forwarding edges};
\end{tikzpicture}
\caption{One transfer of a word by a holder of $C_v$ (\cref{alg:transfer}). The holder applies the share encoder $V_{D,\sigma,p}$ to its at most $J$ registers of the word, padded to $p$, together with $\sigma$ fresh uniform registers. It sends one output along each of its $D$ forwarding edges into the parent committee $C_{v'}$ and retains no old input. As long as at most $\sigma$ outputs reach corrupted holders (shaded), they carry no information about the word.}
\label{fig:transfer}
\end{figure}
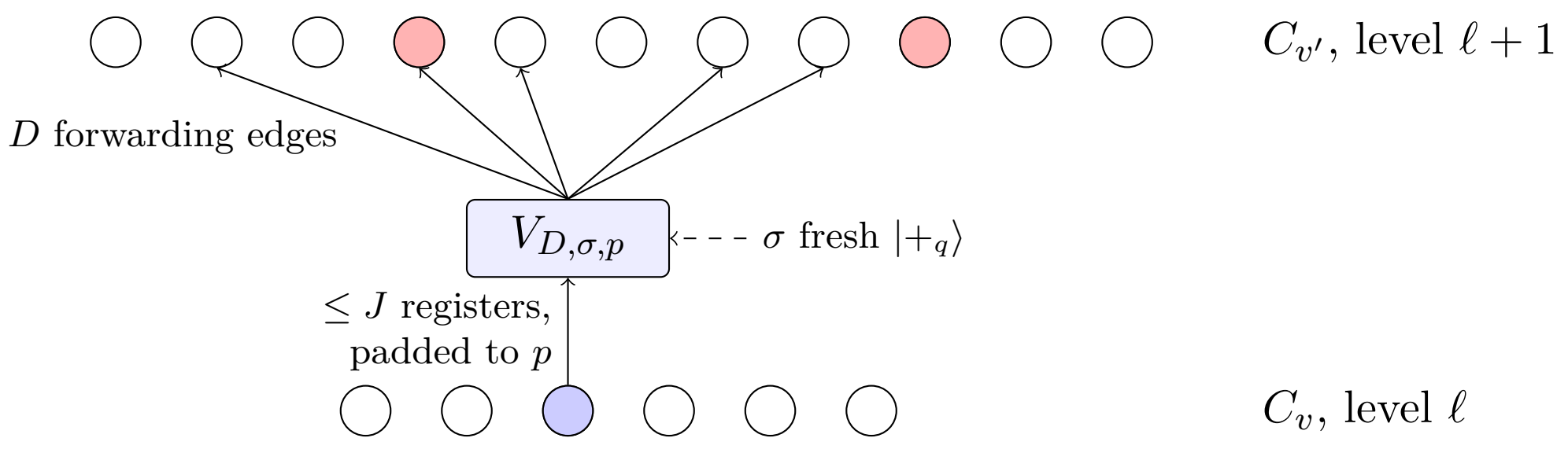

Each holder forwards the unopened words of the arrays in its own election output. If honest members disagree on that output, some may forward an array while others do not. Such arrays are not counted by the survival argument. Nevertheless, the protocol must bound communication on these executions too: a recipient accepts at most one register per authorized array, word, and incoming edge. When consecutive committees are both the full process set, the holders keep their registers and no transfer is needed. The transfer procedure and its conventions are specified in \cref{alg:transfer,sec:transfer-conventions}.

\subsection{Packing and protection under adaptive corruption}\label{sec:ov-protection}

\paragraph{Packing.}
After one transfer, a holder may have up to $J$ registers of a word, one from each incoming neighbor. If it re-encoded them separately, it would send up to $J$ blocks; the next holder could then receive $J^2$ registers. Repeating this would destroy the communication bound.

Instead, each holder encodes all its registers of one word as a single block. Missing inputs are replaced by zero registers, and the input list is padded to a fixed length. Every outgoing edge then carries just one register of that word. Consequently, every recipient again holds at most $J$ registers, regardless of the number of previous transfers. Initially a holder has one register, so $p_1=1$; later we use $p_\ell=J$. Packing changes the representation of a word, not the number of independent words in an array. The condition $D_\ell\ge p_\ell+3\sigma(D_\ell)$ ensures that every block still decodes.

\paragraph{Exposure and protection.}
Privacy of one block does not by itself establish privacy through the whole tree. The adversary can corrupt different holders at different times, and it may seize outgoing messages by corrupting their sender during the transfer round. We therefore follow the complete encoding circuit of each word.

A block is \emph{exposed} if its holder is corrupted by the end of its transfer round. Corrupting the holder later reveals no old input, although the adversary may still obtain new outputs stored there. To account for both direct exposure and later leakage, assign marks backward through the circuit. Mark a currently stored register if its holder is corrupted. Mark a block if it is exposed or if more than its threshold $\sigma$ outputs are marked. Its input wires then count as marked outputs of earlier blocks. A word is \emph{protected} while its dealer block remains unmarked. These marks are used only in the proof; processes do not compute them.

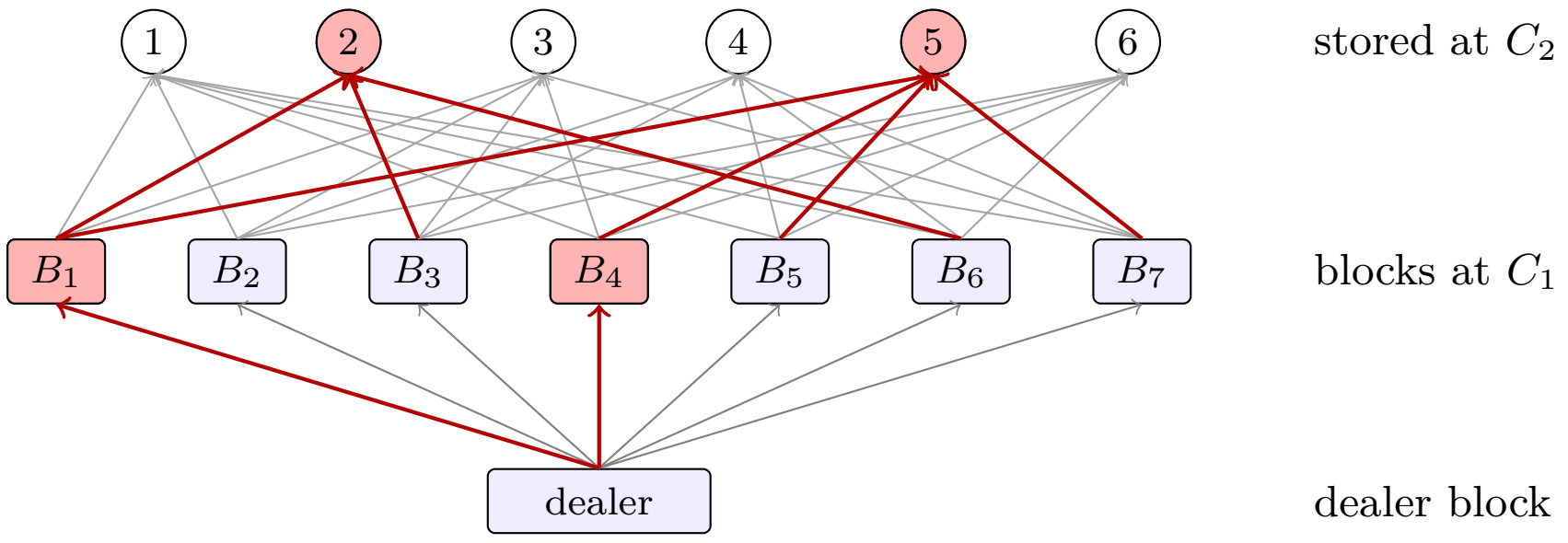
\begin{figure}[t]
\centering
\begin{tikzpicture}[x=1cm,y=1cm,
  st/.style={circle,draw,minimum size=4.6mm,inner sep=0pt,font=\scriptsize},
  stbad/.style={st,fill=red!30},
  bk/.style={draw,rounded corners=1.5pt,minimum width=7mm,minimum height=4.6mm,inner sep=1pt,font=\scriptsize,fill=blue!7},
  bkbad/.style={bk,fill=red!30}]
\foreach \n/\x in {1/1.6,2/3.0,3/4.4,4/5.8,5/7.2,6/8.6}{\node[st] (s\n) at (\x,3.3) {$\n$};}
\node[stbad] at (s2) {$2$}; \node[stbad] at (s5) {$5$};
\foreach \i/\x in {1/0.9,2/2.2,3/3.5,4/4.8,5/6.1,6/7.4,7/8.7}{\node[bk] (u\i) at (\x,1.65) {$B_\i$};}
\node[bkbad] at (u1) {$B_1$}; \node[bkbad] at (u4) {$B_4$};
\node[bk,minimum width=16mm] (dl) at (4.8,0) {dealer};
\foreach \i in {1,...,7}{\draw[->,gray] (dl.north) -- (u\i.south);}
\foreach \i/\list in {1/{1,2,3,5},2/{1,3,4,6},3/{2,3,4,6},4/{1,3,5,6},5/{1,4,5,6},6/{1,2,4,6},7/{1,3,4,5}}{
  \foreach \t in \list {\draw[->,gray!70] (u\i.north) -- (s\t.south);}}
\foreach \i/\t in {1/2,1/5,3/2,4/5,5/5,6/2,7/5}{\draw[->,red!70!black,thick] (u\i.north) -- (s\t.south);}
\foreach \i in {1,4}{\draw[->,red!70!black,thick] (dl.north) -- (u\i.south);}
\node[anchor=west,font=\footnotesize] at (9.8,3.3) {stored at $C_2$};
\node[anchor=west,font=\footnotesize] at (9.8,1.65) {blocks at $C_1$};
\node[anchor=west,font=\footnotesize] at (9.8,0) {dealer block};
\end{tikzpicture}
\caption{The circuit of one word after one transfer, with $\sigma=1$ for the blocks $B_1,\dots,B_7$ at $C_1$ ($D=4$ outputs each) and $\sigma=2$ for the dealer block ($7$ outputs). The holders of the stored registers $2$ and $5$ at $C_2$ are corrupted, so these registers are marked. Block $B_1$ sends two outputs to marked registers, more than $\sigma=1$, and is marked; the holder of $B_4$ was corrupted during its transfer round, so $B_4$ is exposed, hence marked. Every other block at $C_1$ has at most one marked output. The dealer block has two marked outputs, the inputs of $B_1$ and $B_4$, which is within its $\sigma=2$, so it is unmarked and the word is protected. Marked outputs are drawn in red.}
\label{fig:circuit}
\end{figure}

\paragraph{Reconstruction.}
A requester collects measured registers from the current holders, with their incoming-edge labels, and decodes backward through the public circuit. Decoding a latest block recovers its input values; these supply outputs for decoding earlier blocks. Continuing reaches the dealer's input. Former holders need not participate. Figure~\ref{fig:circuit} illustrates how some blocks may be marked while the dealer block remains protected.

\paragraph{Lemma~\ref{lem:disposable} (informal).}
Fix an adversary and an entry history. If the relevant words remain protected on every continuation until their last reconstruction, their unopened values remain jointly uniform and independent of the adversary's information, and every requester honest through reconstruction obtains the corresponding value.

The proof follows the same backward order twice. For privacy, an unmarked block exposes at most $\sigma$ outputs to the remaining adversarial system. The encoder identity replaces these outputs by a fixed state independent of the block's inputs. Eliminating blocks in reverse order eventually removes dependence on the dealer's input. For reconstruction, every unmarked block has enough correct outputs to recover its inputs despite the marked outputs. Repeating the decoding reaches the dealer. A final argument connects these statements to common latent classical values, without assuming that dealers initially sampled classical secrets. The formal proof handles joint attacks and partial openings as well.

\paragraph{Protection along a path.}
It remains to show that arrays counted by the election analysis satisfy this protection condition. The forwarding graphs ensure that any small set of bad recipients receives too many edges from only a small fraction of senders. Thus a bounded bad fraction in the parent committee induces a bounded bad fraction in the child committee. Applying this property backward along a successful array's path bounds the marked outputs of its dealer block.

More precisely, the proof combines holders corrupted by their transfer deadline with holders having too many edges into the next bad set. The sampler bound keeps this combined set below the same fraction at every level, so the error does not accumulate with the path length. Provided that past committees stayed within budget through their transfers and forwarded the array consistently, and the current committee stays within budget, the dealer block remains unmarked. This is the protection result of \cref{lem:path}.

At the start of an election, the relevant past transfers and their corruption deadlines are already fixed. Restricting the current continuation to a budget-respecting adversary therefore makes the protection lemma applicable throughout that election. The Disposable-Share Lemma then supplies the fresh bin words, coin words, and consistent reconstruction needed by the classical analysis. Lemma~\ref{lem:moving-coins} establishes this connection precisely.

The slack below the corruption threshold $1/3$ serves both protection and packing. We reserve part for committee overload and the forwarding graphs, and part for decoding several inputs in one block. Geometrically growing committees let each sender have more outgoing edges than a recipient has incoming edges. This makes room for the packed inputs while preserving the decoding inequality. The exact inequalities are checked together in \cref{app:checks-packed}.

\subsection{Communication and running time}\label{sec:ov-costs}

The construction balances growing committees against a shrinking number of elected arrays. The tree has $h=O(\log n)$ levels and lists of length $w=\Theta_\varepsilon(\log^3n)$. In the packed sharing scheme, each array contains $M=2h+1$ words: one bin word and one coin word per level, and a final word. At each level, the total number of committee memberships is $O_\varepsilon(n\log n)$. Packing bounds each holder's transmission of one unopened word by $O_\varepsilon(\log n)$ registers, each containing $O(\log n)$ qubits.

The resulting quantum cost is $O_\varepsilon(n\log^7n)$ per level and $O_\varepsilon(n\log^8n)$ per epoch; initial distribution costs less. Classical voting is more expensive because committee members communicate pairwise. Summing those costs over the geometrically growing committees gives $O_\varepsilon(n^2\log^8n)$ bits per epoch, including reconstruction and final agreement. Each election takes one constant-round phase per candidate, so the $O(\log n)$ levels require $O_\varepsilon(\log^4n)$ rounds. These bounds hold even in unsuccessful epochs; the detailed accounting is in \cref{app:costs}.

Conditional on any epoch-entry history before the first honest decision, the parameter choices make the probability of that epoch's failing to decide at most $1/n$. Consequently, the expected number of epochs is constant. Including delayed halting gives the expected bounds of \cref{thm:packed}.

\subsection{The lower bound}\label{sec:ov-lower}

To prove \cref{thm:lower}, we follow~\cite{AbrahamCDNPRS19} and split the processes into a set $V$ of about $t/2$ processes and the rest $U$. Consider executions in which all processes of $U$ have the same input, and the processes of $V$ are corrupted and ignore every message from $U$, even its arrival. By validity, the processes of $U$ decide their input, and by no-signalling, the evolution of $V$ does not depend on it. If the expected communication is below about $t^2/16$, some $p\in V$ receives, with probability above $1/2$, at most about $t/2$ messages from $U$, whichever input $U$ has. The adversary can then leave $p$ uncorrupted and instead corrupt the sender of each such message in the round in which it is sent, and suppress the message. Now $p$ is correct, so by agreement it must output the input of $U$ with probability above $1/2$ for both inputs, although its output has the same distribution in both cases. Besides no-signalling, the only quantum ingredient is that suppressing a quantum message acts on the rest of the system exactly as delivering it to a process that discards it, since both apply the partial trace over the message (\cref{sec:lower}).

\section{Protocol specification and randomness requirements}\label{sec:interface}
We specify the tree, the words used by its elections, and the timing and communication rules. We then identify the arrays counted by the analysis and state the randomness guarantees they require. The next section proves agreement from these guarantees; the quantum construction establishes them.

\subsection{Tree, arrays, and committees}\label{sec:sel-spec}

\paragraph{Tree and parameters.} The parameters are integers $b\ge2$ and $w$, both powers of two, the number of levels $h\ge0$, and loss parameters $\beta,\tau,\eta\in(0,1)$. Let $N=wb^h$ be the number of arrays, $r=bw$, $\iota=\lceil\log_2(N+1)\rceil$ and $u=r(\iota+\log_2b)$. Thus $\iota$ bits write an identifier in $\{0,\dots,N\}$, and $u$ bits write $r$ identifier/bin pairs.

The identifiers $1,\dots,N$ form $N/w$ public \emph{level-$0$ lists} $((i-1)w+1,\dots,iw)$. Level $\ell\in[h]$ has $m_\ell=N/(wb^\ell)$ nodes; the same formula gives the number $m_0=N/w$ of level-$0$ lists. Node $v_{\ell,i}$ has the $b$ children $v_{\ell-1,(i-1)b+1},\dots,v_{\ell-1,ib}$, where level-$0$ nodes are the level-$0$ lists, and its \emph{leaf interval} is $I_{v_{\ell,i}}=\{(i-1)wb^\ell+1,\dots,iwb^\ell\}$. Every internal node outputs a list of $w$ identifier slots; each leaf supplies its fixed list, so every election has $r=bw$ input slots and uses $b$ bins. The root is the level-$h$ node; if $h=0$, the root list is the single public level-$0$ list. Identifier $0$ denotes an empty slot.

\paragraph{Arrays and dealers.} Each array has an identifier $x\in[N]$ and contains the words described below. The dealer of array $x$ is $\dealer(x)=1+((x-1)\bmod n)$, so each process deals $\lfloor N/n\rfloor$ or $\lceil N/n\rceil$ arrays. 

\paragraph{Word values.} Let $d=\lceil\log_2(n+1)\rceil$ and $q=2^d>n$. Elements of $\Fq$ are represented by $d$ bits or $d$ qubits in a fixed basis over $\F_2$.

We fix a nonzero $\F_2$-linear map $\psi:\Fq\to\F_2$, which turns the value of a word into a coin bit (\cref{sec:coinphases}). If $S$ is uniform on $\Fq$, its $d$ coordinates are independent uniform bits, and so is $\psi(S)$.

\paragraph{Words and their labels.} We identify a word by $\omega=(x,\lambda)$, where $x$ is its array identifier and $\lambda$ its public label. Its reconstructed value is denoted by $S_\omega$. Each honestly prepared array consists of independent words with public labels: a \emph{bin word} $(\lab{bin},\ell)$ for each level $\ell\in[h]$, a \emph{coin word} $(\lab{coin},\ell)$ for each level, and one \emph{final word} $(\lab{final})$. The bin of a candidate is given by the first $\log_2b$ bits of its bin word. The coin bit of a phase is $\psi$ of the coin word of the phase's provider, and the coin bit of a final phase is $\psi$ of the provider's final word. Bin, coin and final words are distinct words, so an already revealed word never serves as a fresh coin.

Each array therefore has $M=2h+1$ words.

\paragraph{Committees and stage order.} Each internal node $v$ at level $\ell$ represents a public committee $C_v\subseteq P$ of size $k_\ell$. Committees may overlap, and membership is unrelated to dealing. Members run separate local copies of the election. The level loop of \cref{alg:qba} processes levels sequentially, with all committees at a level running concurrently. Let $e_v$ be the boundary at which members produce their local output lists and $d_v$ the end of the delivery round. All nodes at a level have the same delivery deadline. The root committee is $P$; there is no parent delivery, and $e_v=d_v$ is its output deadline. Selection leaves each process's input estimate unchanged.

\subsection{Communication rules and opening times}\label{sec:communication}

\paragraph{Receiving child lists.} Each member of a child committee sends its complete $w$-slot output list to every member of the parent committee. A recipient accepts a well-formed list only if a strict majority of the child's fixed membership sent exactly that list; otherwise it uses $w$ zeros. Level-$0$ lists are public. The accepted lists are concatenated in child order and then \emph{normalized}. A position coming from child $u$ may contain only an identifier in $I_u$, or $0$. The recipient scans the positions in increasing order. A position from child $u$ keeps its identifier only if it lies in $I_u$ and has not occurred at an earlier position; otherwise it becomes $0$. Only kept identifiers count as seen. For identifier/bin tuples, a rejected position becomes $(0,0)$. A faulty child therefore cannot claim, and thereby suppress, another child's identifier.

\paragraph{Opening schedule.} The holders of an array's level-$\ell$ words belong to the committee running its level-$\ell$ election. They open the bin word $(\lab{bin},\ell)$ in that level's bin round. Each member fixes its provider list before the coin phases begin. In phase $j$, a holder opens the coin word $(\lab{coin},\ell)$ of the array in position $j$ of its own list, only after that phase's proposal deadline (\cref{sec:coinphases}). Different local lists may give different requests in a failed election; the authorization rules below still bound the replies. The final word of the array in position $j$ of a holder's root list is opened after the proposal deadline of final phase $j$.

\paragraph{Requests and quotas.} A reconstruction request names one whole word: it carries the node identifier, the array identifier and the word label, not an individual register address. The epoch and round are implicit. Identifier $0$ requires no request. A holder replies once with the measured values of all its stored registers of that word, each paired with its public register address. An address identifies the block that produced the register and its output position; forwarding edges determine these positions. The requester accepts at most one field element for each authorized address held by that sender, ignores extra or unauthorized entries, and replaces missing or malformed entries by $0$. An absent or malformed reply supplies $0$ at every expected address. There is one register per holder after direct distribution and at most $J$ after transfers, so each request has $O(L)$ bits and each reply has an $O(L)$-bit header plus at most $J$ address--value pairs, each of $O(L)$ bits.

A member of $C_v$ may send each holder at most $r$ whole-word bin requests for $v$ and one whole-word coin request per phase; in \FinalAg{} a process may send each holder one whole-word request per final phase. These quotas count words, not registers. Holders check public committee membership and ignore wrong-stage, early, unauthorized and duplicate requests. A request never advances the opening schedule: a holder reports a word only after its scheduled opening under its own frozen provider list, and repeated authorized replies use the same cached readouts. Because each requester has its own quota, faulty requesters cannot use up an honest requester's quota. These rules bound honest replies even in failed elections.

\subsection{Histories, budgets, and counted arrays}

For a boundary $s$, which may lie inside a round, let $B_s$ be the set of processes corrupted by $s$. A \emph{history} $\hist$ is a finite classical execution prefix up to a boundary: all classical states and messages, corruption choices and instrument outcomes so far. Statements conditional on $\hist$ refer to histories of positive probability, for a fixed input vector and a fixed adversarial strategy. Such statements retain the adversary's quantum system $E$, and we write $\rho_{E|\hist}$ for its state conditional on $\hist$. An \emph{epoch} is one iteration of the loop of \cref{alg:qba}; each process still active in that epoch prepares fresh arrays assigned to it.

We condition on past histories, not on whether a committee will stay within its corruption budget in the future. A corruption budget lets us express the required guarantees as bounds on joint events.

A corruption budget bounds the number of faulty committee members through a specified deadline. Since corruption is permanent, a bound at the deadline also bounds every earlier boundary.

\paragraph{Committee budgets.} Set $\gamma=1/3-\varepsilon/2$ and let $a_C=\lfloor(|C|-1)/3\rfloor$ be the voting threshold of committee $C$. For node $v$, write $\Bud_v$ for the event $|C_v\cap B_{d_v}|\le\gamma|C_v|$. This bound must hold through delivery, not merely through local output. For committees of size at least $2/\varepsilon$, $\gamma|C_v|\le a_{C_v}$; the committee assignment below meets this size requirement whenever $C_v\ne P$. For $C_v=P$, the global bound gives $t\le a_P$ and $t\le\gamma n$. The root budget therefore always holds. A committee is \emph{overloaded} when its budget fails.

\paragraph{Scope of the survival analysis.} The good-array and survival analysis below is used only in epochs whose entry history precedes the first honest decision. No honest process has halted at such an entry, so every dealer that remains honest through distribution actually prepares its arrays. Let $s_{\mathrm{dec}}$ be the boundary of the first honest decision, or $\infty$ if no such decision occurs. Randomness guarantees are required only at histories and for reconstructions strictly before $s_{\mathrm{dec}}$. No first honest decision can occur during selection; it occurs only at a final-phase proposal deadline. Once it occurs, \cref{lem:safety} supplies completion independently of all later coins. Later epochs may contain halted, uncorrupted dealers and are covered only by safety and the deterministic cost bounds, not by good-array survival or freshness claims. This restriction is for the analysis; processes need not detect $s_{\mathrm{dec}}$.

\paragraph{Good arrays.} In an epoch with no prior honest decision, let $s_0$ be the end of distribution, and set $\Good=\{x\in[N]:\dealer(x)\notin B_{s_0}\}$. The set $\Good$ is used only in the analysis, and a later corruption of a dealer does not change it.

\begin{definition}[Successful node]\label[definition]{def:success}
Node $v$ is \emph{successful} if $\Bud_v$ holds and all members honest at $e_v$ output the same normalized list $A_v$.
\end{definition}

\paragraph{Counted sets.} The analysis follows the good arrays that survive each level. Let $G_0=\Good$. For $\ell\ge1$, $G_\ell$ consists of the identifiers of $G_{\ell-1}$ that appear in the output of a successful level-$\ell$ node.

Success does not require successful children or honesty beyond the delivery deadline. If $v$ is successful, at least $|C_v|-a_{C_v}>|C_v|/2$ members stay honest through delivery and send $A_v$. Every honest parent member therefore accepts $A_v$, and no other list can reach a majority.

\subsection{Randomness requirements}

\begin{definition}[Fresh common coins]\label[definition]{def:store}
A sharing scheme \emph{provides fresh common coins} if, conditional on every epoch-entry history preceding the first honest decision, the following holds for every level-$\ell$ node $v$, every adversary under which $\Bud_v$ holds on every execution, and every history $\hist_0$ at the common deadline of level $\ell-1$ (the end of distribution if $\ell=1$). Let $W$ be the level-$\ell$ bin and coin words of the arrays of $G_{\ell-1}$ in $v$'s subtree. These words are unopened at $\hist_0$ and have latent values $(S_\omega)_{\omega\in W}$ with two properties:
\begin{enumerate}[label=(\alph*)]
\item At every extending history $\hist$ strictly before $s_{\mathrm{dec}}$, the values of any still-unopened subset $W'\subseteq W$ are jointly uniform and independent of the adversary's quantum memory $E$:
\[
\rho_{S_{W'}E|\hist}=q^{-|W'|}\sum_s\ketbra ss\otimes\rho_{E|\hist}.
\]
\item Every member of $C_v$ that remains honest through a reconstruction of $\omega\in W$ completed strictly before $s_{\mathrm{dec}}$ obtains $S_\omega$.
\end{enumerate}
The same properties, with the same restriction to histories and reconstructions strictly before $s_{\mathrm{dec}}$, hold for the final words of $G_h$, with all processes as requesters, after every selection history in such an epoch and under every adversary. For $h=0$, this history is taken at the end of distribution.
\end{definition}

Only counted words require these guarantees. Other words may be biased or reconstructed inconsistently; requests still finish at their deadlines and return field elements. The guarantees concern a fixed past history and adversaries respecting the budget on every continuation. They do not condition on the original adversary's future corruptions or on the event that no honest decision will occur later; whether a history precedes the first honest decision is determined by that history.

\section{Agreement and selection}\label{sec:classical}
We first analyze the common-coin procedure, then one election and the survival of good arrays through the tree. Finally, we use the selected words for input agreement and bound the number of epochs.

\subsection{Agreement with fresh common coins}\label{sec:coinphases}

We first analyze the common-coin procedure used by the elections and by final agreement. A committee runs \CoinPhases{} on $u$ bits simultaneously. Its phase-$j$ coin comes from the word of the array at position $j$ in a provider list of $r$ identifiers. Each member fixes its own copy of this list before the first phase. The copies may differ, but none changes during the procedure.

Here $k=|C|\ge3a+1$. The estimate has $u$ components; all assignments indexed by $i$ apply to every component in parallel. A single coin bit $c$ is used for all components. Committee elections use $a=a_C$; final agreement uses $k=n$ and $a=t$.

\begin{algorithm}[t]
\caption{\CoinPhases{} (one member of committee $C$)}\label{alg:coinphases}
\KwIn{estimate vector $x$; threshold $a$; frozen providers $(p_1,\ldots,p_r)$}
\KwOut{updated estimate vector $x$}
\For{$j\gets1$ \KwTo $r$}{
  \tcp{Round 1: exchange estimates}
  send $x$ to all members of $C$\nllabel{ln:cp-send}\;
  $y_i\gets v$ if at least $|C|-a$ estimates equal $v$; otherwise $y_i\gets\bot$\nllabel{ln:cp-propose}\;
  \tcp{Round 2: exchange proposals}
  send $y$ to all members of $C$\nllabel{ln:cp-sendprop}\;
  $f_i\gets v$ if at least $a+1$ proposals equal $v$; otherwise $f_i\gets\bot$\nllabel{ln:cp-force}\;
  \tcp{Proposal deadline: the coin remains unopened until this boundary.}
  \tcp{Round 3: request the coin}
  request the level's coin word from provider $p_j$\nllabel{ln:cp-request}\;
  \tcp{Round 4: reconstruct and update}
  $c\gets\psi(\text{reconstructed word value})$\nllabel{ln:cp-coin}\;
  $x_i\gets f_i$ if $f_i\ne\bot$; otherwise $x_i\gets c$\nllabel{ln:cp-adopt}\;
}
\Return{$x$}\;
\end{algorithm}

A member keeps an estimate vector $x\in\{0,1\}^u$, one estimate per component, and in each phase it computes a proposal vector $y$ and a vector $f$ of forced values, both in $\{0,1,\bot\}^u$; $f_i=\bot$ means that nothing is forced for component $i$. 

The voting defaults and empty-provider behavior follow \cref{sec:code-conventions}. The end of round~2 is the proposal deadline. The opening schedule of \cref{sec:communication} identifies the coin word for each frozen provider, and $\psi$ extracts its coin bit.

There is no early halting: every member runs all $r$ phases and then outputs its estimate vector.

\begin{lemma}[One good phase]\label{lem:phase}
Consider one phase of \CoinPhases{} and one component. While at most $a$ members are corrupted, three things hold: (i) no two honest members propose different bits for the component; (ii) no two honest members record different forced bits for it; and (iii) if all honest estimates for the component equal $v$ at the start of the phase, then every member honest at the end of the phase has estimate $v$.

Moreover, fix a history $\hist$ at the proposal deadline of the phase, and suppose that the coin bit of the phase is decoded identically by all members honest through decoding and is uniform conditioned on $\hist$. Let $\mathsf{Comp}$ be the event that at most $a$ members are corrupted through the end of the phase, and $\mathsf{Split}$ the event that the honest estimates for the component are not unanimous at the end of the phase. Then $\Pr[\mathsf{Comp}\wedge\mathsf{Split}\mid\hist]\le\tfrac12$.
\end{lemma}

\begin{proof}[Proof of \cref{lem:phase}]
The threshold rules have three consequences: honest proposals cannot conflict, every forced bit has an honest supporter, and unanimity persists regardless of the coin. Once these are established, a fresh common coin aligns the remaining estimates with probability at least one half.

Fix the component and suppose that at most $a$ members are corrupted. Recall that $k\ge3a+1$. Suppose that two honest members propose opposite values in line~\ref{ln:cp-propose}. Each of them received at least $k-a$ supporting estimates, so the two sets of senders intersect in at least $k-2a\ge a+1$ identities. At most $a$ of these are corrupted by the end of the round, so the intersection contains a sender that stayed honest through the round. Its estimate, sent in line~\ref{ln:cp-send}, was delivered unchanged to everyone, so it cannot support opposite values. This gives~(i).

A forced value recorded in line~\ref{ln:cp-force} has at least $a+1$ supporting proposals, and hence a proposer that stayed honest through the round; by~(i), honest forced values agree.

For~(iii), suppose that all honest estimates equal $v$. The at least $k-a$ members that stay honest through round~1 send $v$ to everyone. Every honest member proposes $v$, since no honest member can collect $k-a>a$ votes for $1-v$. Every honest member then receives at least $k-a\ge a+1$ proposals of $v$, and at most $a$ of $1-v$, so it is forced to $v$ in line~\ref{ln:cp-force} and keeps $v$ in line~\ref{ln:cp-adopt}, regardless of the coin.

Now consider the second part, and fix $\hist$. If more than $a$ members are already corrupted at $\hist$, then $\mathsf{Comp}$ fails. Otherwise, by~(ii), the forced values that honest members have recorded at $\hist$ take at most one value $v$. If some honest member is forced to $v$, then a coin equal to $v$ aligns every surviving honest member: the forced ones keep $v$, and the others adopt the coin, which they decode identically. This has probability $1/2$, because the coin is uniform given $\hist$ and $v$ is determined by $\hist$. If nobody is forced, either value of the coin aligns them. Later corruptions can only remove members; they cannot change forced values that survivors have already recorded. Thus $\Pr[\mathsf{Comp}\wedge\mathsf{Split}\mid\hist]\le\tfrac12$.
\end{proof}

Some providers may supply biased coins or different decoded values to different members. Nevertheless, enough fresh common coins ensure agreement. The members do not need to know which providers supply them.

\begin{lemma}[Enough good phases]\label{lem:enough}
Fix a history $\hist_0$ no later than the start of the procedure. Suppose that for $g$ phases $j_1<\dots<j_g$, conditioned on any history at the proposal deadline of phase $j_i$, the coin bit of the phase is decoded identically by all members honest through decoding and is uniform. Let $\mathsf{Comp}$ be the event that at most $a$ members are corrupted through the output, and $\mathsf{Diff}$ the event that two honest members output different vectors. Then
\begin{equation}\label{eq:enough}
\Pr\big[\mathsf{Comp}\wedge\mathsf{Diff}\ \big|\ \hist_0\big]\le u\,2^{-g}.
\end{equation}
While at most $a$ members are corrupted, every component that is unanimous among honest members at the start stays unchanged.
\end{lemma}

\begin{proof}[Proof of \cref{lem:enough}]
For one component, unanimity persists while the budget holds, and each fresh common-coin phase halves the remaining probability of disagreement. No independence between phases is needed.

Fix a component. Let $U_0$ be the whole conditional sample space, and let $U_i$ be the event that after phase $j_i$ at most $a$ members are corrupted and the honest estimates are not unanimous. Persistence of unanimity gives $U_i\subseteq U_{i-1}$. The history $\hist_i$ at the proposal deadline of phase $j_i$ determines $U_{i-1}$. By \cref{lem:phase},
\[
\E[\one_{U_i}\mid\hist_i]\le\tfrac12\one_{U_{i-1}},
\qquad
\Pr[U_i\mid\hist_0]\le\tfrac12\Pr[U_{i-1}\mid\hist_0].
\]
The second inequality follows by conditional expectation. Iterating gives $\Pr[U_g\mid\hist_0]\le2^{-g}$.

If the budget holds through output but the output vectors differ, some component is still nonunanimous. By persistence, its event $U_g$ must have occurred. A union bound over the $u$ components proves \eqref{eq:enough}. The components may all use the same coin: this last step requires no independence between them. Persistence also proves the final claim.
\end{proof}
\subsection{One committee election}\label{sec:sel-alg}\label{sec:sel-analysis}
Consider one committee election. It uses the opening times, normalization, and request limits of \cref{sec:communication}. The parameters remain symbolic until \cref{sec:proof-packed}.

List acceptance and normalization use \cref{sec:communication}; at level~1 the public leaf lists are used directly. A bin is the first $\log_2 b$ bits of its reconstructed word. Empty positions belong to bin~$0$, and ties between lightest bins are resolved by the smallest bin index. Delivery sends the complete local list to every parent member; at the root it keeps the list for final agreement. The common-coin call runs $r$ phases on the $u$ bits of all pairs.

\begin{algorithm}[t]
\caption{\LBE{} (one member of committee $C_v$)}\label{alg:election}
\KwIn{the $b$ child lists, each with $w$ positions}
\KwOut{a local list $A$ of $w$ array identifiers}
accept child lists, concatenate them, and normalize to $Y=(y_1,\ldots,y_r)$\nllabel{ln:sel-receive}\;
freeze $Y$ as the provider list for all coin phases\nllabel{ln:sel-freeze}\;
reconstruct each candidate's level-$\ell$ bin word and extract its bin $z_j$\nllabel{ln:sel-bins}\;
run \CoinPhases{} on the identifier/bin pairs, with threshold $a_{C_v}$ and providers $Y$\nllabel{ln:sel-agree}\;
normalize the resulting pairs $(y'_j,z'_j)$\nllabel{ln:sel-renorm}\;
$S\gets$ positions assigned to a lightest bin\nllabel{ln:sel-lightest}\;
$T\gets$ the first $w-|S|$ remaining positions\nllabel{ln:sel-fill}\;
$A\gets$ identifiers at $S$, followed by those at $T$, in increasing position order within each part\nllabel{ln:sel-output}\;
deliver $A$\nllabel{ln:sel-deliver}\;
\end{algorithm}

\paragraph{Provider list and output.} The provider list is fixed before the coin phases because estimates change during agreement, whereas the provider for each phase must already be determined. Since $r$ positions are assigned to $b$ bins, the lightest bin has $|S|\le r/b=w$ positions. Filling the output to length $w$ preserves all these positions. The chosen order matters only insofar as it is a deterministic function of the agreed tuple vector. In \QBAP{}, holders transfer the selected arrays' unopened words during delivery.

\paragraph{Counted inputs.} For a level-$\ell$ node $v$, a \emph{counted input} is a position holding an identifier of $G_{\ell-1}$. Let $g_v$ be the number of counted inputs, and fix the \emph{entry history}, the history at the common deadline of level $\ell-1$ (the end of distribution if $\ell=1$). It precedes the opening of the level's bin words, and it determines $G_{\ell-1}$ and $g_v$. Counted identifiers come from successful children, so every honest member has the same identifier at each counted position. Normalization cannot remove a counted identifier: the positions of a successful child are common and distinct, and no other child can place an identifier from that child's interval.

\begin{lemma}[One election]\label{lem:election}
Assume the sharing scheme provides fresh common coins, and consider an epoch entered before the first honest decision. Let $v$ be a level-$\ell$ node with $g=g_v\ge\tau r$ counted inputs. Let $\mathsf{Fail}_v$ be the event that the members of $v$ honest at $e_v$ output different lists, or that fewer than $(g/r-\eta)w$ entries of the common output lie in $G_{\ell-1}$. Conditional on $v$'s entry history,
\begin{equation}\label{eq:xi}
\Pr[\Bud_v\wedge\mathsf{Fail}_v]\ \le\ \xi:=u\,2^{-\tau r}+b\,e^{-\eta^2w/2}.
\end{equation}
\end{lemma}

\begin{proof}[Proof of \cref{lem:election}]
There are two possible failures: the members do not agree on the tuples, or some bin contains too few counted tuples. If neither occurs, even a bin chosen using adversarial tuples contains enough counted identifiers, and filling the output removes none of them. We bound these two events after reducing to a continuation that always respects the committee budget.

This reduction is necessary because conditioning on the original adversary's future compliance could bias the coins. Fix the entry history $\hist_0$. If $\Bud_v$ has already been violated, the joint event in the statement is impossible. Otherwise, start a modified continuation $A'_{\hist_0}$ from the same conditional state of all systems as the original adversary $A$, including its quantum memory. This continuation follows $A$ until a corruption would violate $\Bud_v$. It refuses that corruption and thereafter follows any fixed strategy that respects both the committee budget and the global corruption bound. On every branch where $A$ respects the committee budget, no refusal occurs and both continuations apply the same operations. Summing the probabilities of these branches gives
\[
\Pr_A[\mathsf{Fail}_v\wedge\Bud_v\mid\hist_0]
\le \Pr_{A'_{\hist_0}}[\mathsf{Fail}_v\mid\hist_0].
\]
Under the modified continuation the committee always remains within its voting threshold, since $\gamma|C_v|\le a_{C_v}$, and the randomness interface applies from $\hist_0$. We bound disagreement and a deficient bin under this continuation, then show that outside these events the election succeeds.

\emph{Agreement.} The $g$ counted positions name arrays of $G_{\ell-1}$. Their coin words for this level are unopened at the respective proposal deadlines and, being fresh common coins, are decoded identically by honest members and are uniform given any history at those deadlines. We apply \cref{lem:enough} to line~\ref{ln:sel-agree} conditionally on the history after the bin stage, and then average. The agreed vectors differ with probability at most $u2^{-g}\le u2^{-\tau r}$. On the complementary event, all honest members hold the same tuple vector, and every counted tuple is unchanged: its identifier and bin are reconstructed identically, so its bits start unanimous. The identifiers of all positions from successful children also stay unchanged, so the normalization in line~\ref{ln:sel-renorm} keeps every counted identifier.

\emph{Bins.} The counted identifiers are distinct, and their bin words are jointly uniform and independent of the entry history, so their bins are independent and uniform over the $b$ bins. Fix a bin. Its counted occupancy $X$ is binomial with mean $\mu=g/b\le w$, and the target occupancy is $\mu-\eta w$. If this target is nonpositive, deficiency is impossible. Otherwise \cref{lem:chernoff} gives
\[
\Pr[X<\mu-\eta w]
\le \exp\!\left(-\frac{\eta^2w^2}{2\mu}\right)
\le e^{-\eta^2w/2}.
\] A union bound over the $b$ bins bounds the probability that some bin is deficient by $b\,e^{-\eta^2w/2}$.

\emph{Selection.} Outside these two events, every bin contains at least $(g/r-\eta)w$ counted tuples, however the other tuples were chosen, so the bin selected in line~\ref{ln:sel-lightest} does too. All its positions fit into the $w$ output slots, and filling in line~\ref{ln:sel-fill} removes none of them. Since the output is a function of the common tuple vector, all honest members output the same list. Hence $\mathsf{Fail}_v$ does not occur outside the two events, and \eqref{eq:xi} follows.
\end{proof}

\subsection{Selection through the tree}\label{sec:selection}

All survival and root-list claims in this subsection concern epochs entered before the first honest decision. The committee assignment itself must work for every possible fault set, since the adversary sees the assignment before choosing corruptions. The next lemma bounds both the number of overloaded committees and the memberships of each process.

\begin{lemma}[Committees; proof in \cref{app:committees}]\label{lem:committees}
Let $\beta\in(0,1]$ and let $\gamma=1/3-\varepsilon/2$. Consider a level with $m$ nodes and committee size $k$, where either $k=n$ or
\[
k\ \ge\ K(m):=\frac{2}{\varepsilon^2}\Big(\frac{n\ln2+2}{\max\{1,\beta m\}}+\ln\frac e\beta\Big).
\]
There is a deterministic public assignment of committees to the level's nodes with two properties. First, for every $F\subseteq P$ with $|F|\le t$, at most $\beta m$ committees $C$ have $|C\cap F|>\gamma|C|$. Second, every process lies in at most $\Delta=\lceil\max\{2emk/n,\,3L\}\rceil$ of the level's committees.
\end{lemma}

All processes use the first valid assignment in a fixed public order. Its existence and the exhaustive search are described in \cref{app:committees}.

\paragraph{Counting across levels.} Define $F_0=|\Good|/N$ and $F_\ell=|G_\ell|/(m_\ell w)$ for $\ell\in[h]$. An unsuccessful node contributes nothing but keeps its $w$ slots in the denominator. A node with $g_v<\tau r$ is \emph{sparse}. Let $\mathcal E$ be the event that $\Bud_v\Rightarrow\neg\mathsf{Fail}_v$ for every node $v$ with $g_v\ge\tau r$. The entry history determines whether $g_v\ge\tau r$, so \cref{lem:election} and a union bound give
\begin{equation}\label{eq:delta}
\Pr[\neg\mathcal E]\le\delta:=\xi\sum_{\ell=1}^h m_\ell ,
\end{equation}
also conditionally on any epoch-entry history before an honest decision.

\begin{lemma}[Level induction]\label{lem:induction}
Suppose that, for the realized fault set at each level's common deadline, at most $\beta m_\ell$ level-$\ell$ budgets fail. Then on $\mathcal E$, $F_\ell\ge F_{\ell-1}-\beta-\tau-\eta$ for every $\ell\in[h]$.
\end{lemma}

\begin{proof}[Proof of \cref{lem:induction}]
There are three losses: overloaded nodes discard at most a $\beta$ fraction of the available input slots, sparse nodes discard at most a $\tau$ fraction, and the remaining elections lose at most $\eta$ in output density.

Write $m=m_\ell$. Every counted identifier reaches its parent intact, so
\[
\sum_{v\text{ at level }\ell}g_v
=|G_{\ell-1}|=mrF_{\ell-1}.
\]
Removing the at most $\beta m$ overloaded nodes loses at most $\beta mr$ counted inputs. Removing sparse nodes loses at most $\tau mr$ more. On $\mathcal E$, each remaining node is successful and retains at least $w(g_v/r-\eta)$ counted identifiers. Thus
\[
|G_\ell|\ge \frac wr\bigl(mrF_{\ell-1}-\beta mr-\tau mr\bigr)-\eta mw
=mw\bigl(F_{\ell-1}-\beta-\tau-\eta\bigr).
\]
Division by $mw$ proves the claim.
\end{proof}

The committee assignment guarantees the hypothesis: at a level's common deadline, apply its fault-set bound to the processes corrupted by that deadline. A budget fails precisely when this set exceeds the committee's allowed fraction.

\begin{proposition}[Root list]\label[proposition]{prop:root}
Suppose the hypothesis of \cref{lem:induction} holds and $F_0-h(\beta+\tau+\eta)>1/2$. Then on $\mathcal E$ the root is successful, its common list contains more than $w/2$ identifiers of $G_h$, and, if the sharing scheme provides fresh common coins, their final words are jointly uniform and independent of the selection history and the adversary's system.
\end{proposition}

\begin{proof}[Proof of \cref{prop:root}]
If $h=0$, the root list is public and contains $|G_0|=wF_0>w/2$ good identifiers. For $h\ge1$, \cref{lem:induction} gives $F_h>1/2$. There is only one root; if it were unsuccessful, $G_h$ would be empty and $F_h=0$. It is therefore successful, with $|G_h|=wF_h>w/2$. In either case the final words remain unopened, so their joint freshness follows from \cref{def:store}.
\end{proof}

Since each process deals at most $\lceil N/n\rceil\le N/n+1$ arrays, $F_0\ge1-t/n-t/N\ge\frac23+\varepsilon-t/N$.

\subsection{Final agreement and repetition}

For phase $j\le w$, the provider is array $x_j$ of the local root list. The extra phase $w+1$ has the empty provider and therefore coin $0$. The decision state is persistent: a process that decided in the previous epoch still executes the next final phase before halting. The flag $\mathit{haltAfterPhase}$ records this state at phase entry, so deciding during the current phase never causes an immediate halt.

\begin{algorithm}[t]
\caption{\FinalAg{} (process $p$, one epoch)}\label{alg:finalagreement}
\KwIn{estimate $e_p$; local root list $A=(x_1,\ldots,x_w)$; persistent decision state}
\KwOut{updated estimate, unless $p$ halts}
$x\gets e_p$\;
\For{$j\gets1$ \KwTo $w+1$}{
  $\mathit{haltAfterPhase}\gets$ whether $p$ has already decided\;
  \tcp{Round 1: exchange estimates}
  send $x$ to all processes\nllabel{ln:sel-fa-send}\;
  $y\gets v$ if at least $n-t$ estimates equal $v$; otherwise $y\gets\bot$\nllabel{ln:sel-fa-propose}\;
  \tcp{Round 2: exchange proposals}
  send $y$ to all processes\nllabel{ln:sel-fa-sendprop}\;
  \lIf{$p$ is undecided and at least $n-t$ proposals equal $v$}{decide $v$\nllabel{ln:sel-fa-decide}}
  $f\gets v$ if at least $t+1$ proposals equal $v$; otherwise $f\gets\bot$\nllabel{ln:sel-fa-force}\;
  \tcp{Proposal deadline: only now may the phase's word be opened.}
  \tcp{Round 3: request the coin}
  request the final word from the phase's provider\;
  \tcp{Round 4: reconstruct and update}
  $c\gets\psi(\text{reconstructed word value})$\nllabel{ln:sel-fa-coin}\;
  $x\gets f$ if $f\ne\bot$; otherwise $x\gets c$\nllabel{ln:sel-fa-adopt}\;
  \lIf{$\mathit{haltAfterPhase}$}{halt\nllabel{ln:sel-fa-halt}}
}
\Return{$x$}\;
\end{algorithm}

\paragraph{Agreement among all processes.} Each process keeps its input as a binary estimate, separate from all committee estimates. \FinalAg{} (\cref{alg:finalagreement}) runs $w+1$ phases of \CoinPhases{}, called \emph{final phases}, among all processes, with $k=n$, $a=t$ and the conventions of \CoinPhases{}. The coin of phase $j\le w$ comes from the final word of the $j$th root identifier, and phase $w+1$ uses coin $0$. We add one rule: at the proposal deadline, a process \emph{decides} $v$ if at least $n-t$ received proposals equal $v$ (line~\ref{ln:sel-fa-decide}). A process keeps updating its estimate after deciding and remains active through the end of the next final phase, including any selection rounds in between if an epoch boundary occurs; then it halts (line~\ref{ln:sel-fa-halt}). Undecided processes start a fresh epoch with their current estimates. The schedule is public and never waits for a common decision.

\paragraph{Safety and progress.}

\begin{lemma}[Safety for any coins]\label{lem:safety}
Agreement and validity hold whatever coin values the processes decode. After the first honest decision, every correct process decides the same value by the end of the next final phase.
\end{lemma}

\begin{proof}[Proof of \cref{lem:safety}]
A first decision forces a common estimate in its own phase; the next final phase turns that common estimate into decisions everywhere. Delayed halting ensures that the processes needed for this second phase are still active.

Before the first honest decision, no honest process has halted, so \cref{lem:phase} applies with $n>3t$. Suppose the first honest decision is $v$. Its $n-t$ supporting proposals include at least $n-2t>t$ from processes honest through the proposal round. These proposals reach every process unchanged. Every surviving honest process is therefore forced to $v$ and adopts it, regardless of its coin. Honest proposals cannot conflict, so no honest process decides $1-v$ in that phase.

All correct processes remain active through the next final phase, including any intervening selection rounds. They enter that phase with estimate $v$, send at least $n-t$ estimates and then proposals of $v$, and all decide $v$. This proves agreement and the stated decision deadline.

For validity, suppose all correct inputs equal $v$. Selection does not change input estimates. In the first final phase, the at least $n-t$ correct processes send estimate $v$, then propose $v$, and every correct process decides $v$.
\end{proof}

\begin{lemma}[Progress per epoch]\label{lem:progress}
Assume the sharing scheme provides fresh common coins and the hypotheses of \cref{prop:root} hold. Conditional on any epoch-entry history before the first honest decision, the probability that the epoch ends with some correct process undecided is at most $p_n:=\delta+2^{-w/2}$.
\end{lemma}

\begin{proof}[Proof of \cref{lem:progress}]
Fix an epoch-entry history before the first honest decision. Selection fails to produce the root list of \cref{prop:root} with probability at most $\delta$. Fix any selection history outside that failure event. The root list has more than $w/2$ counted positions. At a counted phase whose proposal deadline precedes the first honest decision, its final word is fresh by \cref{def:store} and reconstructs consistently: there can be no first honest decision between that deadline and the end of reconstruction, since decisions occur only at proposal deadlines. If a first honest decision has already occurred, \cref{lem:safety} applies instead.

Until the first honest decision, all correct processes remain active. The halving argument of \cref{lem:enough}, applied to the single input estimate, bounds the probability of passing these counted phases without either alignment or an honest decision by $2^{-w/2}$. If alignment occurs in phase $j\le w$, every correct process decides in phase $j+1$. If an honest decision occurs instead, \cref{lem:safety} gives the same deadline. The extra phase ensures $j+1\le w+1$ even when the last useful coin is in phase $w$.

Averaging over the successful selection histories and adding the selection failure probability gives $p_n=\delta+2^{-w/2}$.
\end{proof}

\begin{corollary}[From a sharing scheme to agreement]\label[corollary]{cor:generic}
Suppose a sharing scheme provides fresh common coins, the committees satisfy \cref{lem:committees} at every level with corruption fraction $\gamma$, and $F_0-h(\beta+\tau+\eta)>1/2$ holds in every epoch entered before the first honest decision. Then agreement and validity hold on every execution. If moreover $p_n=\delta+2^{-w/2}<1$, termination holds with probability~1, the expected number of epochs through the last correct halt is at most $(1-p_n)^{-1}+2$, and the expected cost is at most this number times any deterministic per-epoch cost bound.
\end{corollary}

\begin{proof}[Proof of \cref{cor:generic}]
Agreement and validity follow from \cref{lem:safety}. A uniform conditional bound on epoch failure gives a geometric tail, even when the adversary correlates different epochs. For termination, \cref{lem:committees} supplies the hypothesis of \cref{lem:induction} in every epoch entered before the first honest decision, so with the root inequality the hypotheses of \cref{prop:root} hold. Fresh preparation in line~\ref{ln:sel-qba-deal} makes \cref{def:store} apply up to the first honest decision in each such epoch, even though the adversary may retain registers from earlier epochs; they are part of its system $E$. Let $T$ be the epoch of the first honest decision. Applying \cref{lem:progress} at successive epoch entries gives
\[
\Pr[T>i]\le p_n^i,\qquad
\E[T]=\sum_{i\ge0}\Pr[T>i]\le\frac1{1-p_n}.
\]
Thus $\Pr[T=\infty]=0$. After the first decision, one further final phase suffices for all correct decisions, and one more covers their delayed halts. Allowing at most two further epochs therefore covers all correct halts. Since the per-epoch cost bounds are deterministic, the expected cost is at most the expected number of epochs times such a bound.
\end{proof}

\section{Implementing the interface with disposable shares}\label{sec:moving}
We define the share encoder and the packed transfer procedure before proving their guarantees. We then view the transfers as circuits, prove privacy and reconstruction for protected words, and show that arrays surviving the elections follow protected paths.

\subsection{Share encoder}

All evaluation points are distinct nonzero elements of $\Fq$. Since $q-1\ge n$, there are enough points for every block. Polynomial reconstruction uses the following decoding convention.

Polynomial evaluation codes, their minimum distance and unique decoding are recalled in \cref{app:algebra} (\cref{lem:distance}). These are Reed--Solomon codes~\cite{ReedSolomon60}, and their connection with polynomial secret sharing~\cite{Shamir79} was observed in~\cite{McElieceSarwate81}. For a polynomial-degree bound $0\le d_{\mathrm{poly}}<D$ and $y\in\Fq^D$, let $\Dec_{d_{\mathrm{poly}}}(y)$ denote unique decoding to a polynomial of degree at most $d_{\mathrm{poly}}$, within radius $\lfloor(D-d_{\mathrm{poly}}-1)/2\rfloor$. The degree bound is $t$ in the basic sharing example and $p+\sigma-1$ for a transfer block. A decoder always returns a value: when no codeword lies within the decoding radius, it returns $0$ (or an all-zero vector).

We begin with the encoder used at distribution and at every transfer. For integers $p\ge1$, $\sigma\ge0$, $p+\sigma\le D\le n$, and distinct nonzero points $\alpha_1,\dots,\alpha_D\in\Fq$, define
\begin{equation}\label{eq:enc}
V_{D,\sigma,p}\ket{\mathbf s}=q^{-\sigma/2}\sum_{\mathbf r\in\Fq^\sigma}\ \bigotimes_{i=1}^D\ket{f_{\mathbf s,\mathbf r}(\alpha_i)},\qquad f_{\mathbf s,\mathbf r}(X)=\sum_{j=0}^{p-1}s_jX^j+\sum_{j=0}^{\sigma-1}r_jX^{p+j}.
\end{equation}
We call it a \emph{packed} share encoder when $p>1$. Here $\mathbf s=(s_0,\dots,s_{p-1})$ is a basis label, not a vector that the holder learns, and the inputs may be entangled with other systems. The encoder is implemented by an $\F_2$-linear unitary acting on the input and auxiliary registers: a Vandermonde transformation of the $p$ inputs, $\sigma$ fresh $\ket{+_q}$ registers and $D-p-\sigma$ zeros, realizable by CNOT gates without workspace. All $D$ resulting registers are outputs. The encoder therefore \emph{consumes} its inputs and keeps no copy.

The quantum-sharing example of \cref{sec:ov-storage} is the special case
\begin{equation}\label{eq:share}
V_{n,t,1}\ket{+_q}=q^{-(t+1)/2}\sum_{f:\,\deg f\le t}\ \bigotimes_{i=1}^n\ket{f(\alpha_i)} .
\end{equation}

\begin{lemma}[Share encoder]\label{lem:encoder}
\leavevmode
\begin{enumerate}[label=(\alph*)]
\item \emph{(Privacy.)} If $D\ge p+2\sigma$, then for every $A\subseteq[D]$ with $|A|\le\sigma$ and every operator $X$ on the inputs,
\begin{equation}\label{eq:encoder-privacy}
\Tr_{\bar A}\big(V_{D,\sigma,p}XV_{D,\sigma,p}^\dagger\big)=\Tr(X)\,q^{-|A|}I_A .
\end{equation}
\item \emph{(Decoding.)} If $D\ge p+3\sigma$, computational-basis readout followed by polynomial decoding recovers all $p$ input coefficients despite $\sigma$ erroneous or missing reports.
\end{enumerate}
\end{lemma}

\begin{proof}[Proof of \cref{lem:encoder}]
Privacy follows from two facts: tracing out the other outputs removes off-diagonal input terms, and the random polynomial coefficients make the remaining diagonal terms uniform. Decoding then follows from the distance of the evaluation code.

\emph{Off-diagonal terms.} Both sides of \eqref{eq:encoder-privacy} are linear in $X$, so it suffices to consider a matrix unit $X=\ketbra{\mathbf s}{\mathbf s'}$. By \eqref{eq:enc}, $V_{D,\sigma,p}XV_{D,\sigma,p}^\dagger$ is $q^{-\sigma}$ times the sum, over $\mathbf r,\mathbf r'\in\Fq^\sigma$, of $\bigotimes_{i}\ketbra{f_{\mathbf s,\mathbf r}(\alpha_i)}{f_{\mathbf s',\mathbf r'}(\alpha_i)}$. Tracing out the registers outside $A$ keeps exactly the terms in which $f_{\mathbf s,\mathbf r}$ and $f_{\mathbf s',\mathbf r'}$ agree at all $D-|A|\ge p+\sigma$ points outside $A$. Both polynomials have degree at most $p+\sigma-1$, so they are then identical; hence $\mathbf s=\mathbf s'$ and $\mathbf r=\mathbf r'$, and the partial trace vanishes when $\mathbf s\ne\mathbf s'$, as does $\Tr(X)$. 

\emph{Diagonal terms.} For $\mathbf s=\mathbf s'$ the partial trace equals $q^{-\sigma}\sum_{\mathbf r}\ketbra{y_{\mathbf r}}{y_{\mathbf r}}$, where $y_{\mathbf r}=(f_{\mathbf s,\mathbf r}(\alpha_i))_{i\in A}$. The random part of $y_{\mathbf r}$ is $\mathbf r$ multiplied by an $|A|\times\sigma$ Vandermonde matrix with the nonzero row factors $\alpha_i^p$, which has full row rank since $|A|\le\sigma$. Every evaluation vector on $A$ is therefore equal to $y_{\mathbf r}$ for exactly $q^{\sigma-|A|}$ values of $\mathbf r$, and the partial trace is $q^{-|A|}I_A=\Tr(X)\,q^{-|A|}I_A$.

\emph{Decoding.} The outputs are the evaluations of a polynomial of degree at most $p+\sigma-1$ at $D$ distinct points. This code has distance $D-(p+\sigma-1)\ge2\sigma+1$, so unique decoding corrects $\sigma$ errors (\cref{lem:distance}), where a missing report counts as an error. It recovers the whole polynomial, and in particular its $p$ input coefficients.
\end{proof}

Identity \eqref{eq:encoder-privacy} says that the channel from the inputs to any $\sigma$ outputs is a \emph{replacement channel}: it discards the inputs and prepares a fixed state. Since it is an identity of linear maps, it holds also in the presence of arbitrary reference systems. Every encoder below satisfies $D\ge p+3\sigma$.

\subsection{Packed transfers}
\QBAP{} is \QBA{} (\cref{alg:qba}) with the sharing scheme of \cref{sec:ov-reduce}: each word is dealt to the committee that first elects its array, and it moves up with the array by \cref{alg:transfer}. Coin words are opened at the times specified in \cref{sec:communication}, and each array has $M=2h+1$ words. Budgets use the corruption fraction $\gamma$. For an encoder with $m$ outputs, the threshold is $\sigma(m)$, where
\[
\gamma=\tfrac13-\tfrac\varepsilon2,\qquad \sigma(m)=\lfloor\theta m\rfloor,\quad\theta=\tfrac13-\tfrac\varepsilon4 .
\]
For committees of size at least $2/\varepsilon$ we have $\gamma|C|\le a_C$, so a committee within its budget is also within the voting threshold of \CoinPhases{}. For a level $\ell$ from which arrays are transferred, i.e., with $k_\ell<n$, every member of a level-$\ell$ committee has $D_\ell$ forwarding edges. A block applied at level $\ell$ has $p_\ell$ inputs, where $p_1=1$ and $p_\ell=J$ for $\ell\ge2$; $J$ is a uniform bound on the in-degree of every forwarding graph. For now the graphs and capacities are parameters; \cref{sec:proof-packed} chooses them and verifies $D_\ell\ge p_\ell+3\sigma(D_\ell)$. Encoders act on one word at a time, a word being identified by its array identifier and word label.

The initial distribution specified in \cref{sec:transfer-conventions} deals each word to the level-$1$ committee above its array's level-$0$ list, or to $P$ if $h=0$; if that committee is $P$, this is single-block sharing with threshold $\sigma(n)\ge t$ (\cref{cor:single-block}). If both $C_v$ and $C_{v'}$ equal $P$, there is no transfer, and the registers stay where they are (\cref{sec:transfer-conventions}). Different word labels are never packed together, so opening one word never measures a block that contains another. A recipient accepts at most one register per authorized array, word and in-edge (\cref{sec:transfer-conventions}), and each sender forwards at most its own $w$ selected arrays (line~\ref{ln:dt-own}), even when local outputs disagree. Each array follows one path of committees, and a holder at level $\ell\ge2$ receives its registers of a word from at most $J$ in-neighbors in the unique preceding committee. So it has at most $J$ registers of a word, one per in-edge, and they fit into the list of line~\ref{ln:dt-pad}. This holds even when elections fail, since the acceptance rule in \cref{sec:transfer-conventions} does not depend on whether the election succeeded.

\subsection{Protection and reconstruction}

A word passes through its dealer's encoder and then through one block at each holder that transfers it. We represent these operations as a circuit. This lets us track both what the adversary can learn and what a requester can reconstruct, even when corruptions occur between transfers.

Fix one word. Its \emph{dealer block} applies $V_{D,\sigma,1}$ to a fresh $\ket{+_q}$ and sends output $i$ to the $i$th member of a public holder list. A \emph{transfer block} is applied by a holder in some round: the holder applies an encoder $V_{D,\sigma,p}$ to $p$ of its registers of the word, padded with zero registers, and sends the $D$ outputs to $D$ distinct holders. One of them may be the holder itself, in which case it keeps that output as a new register. Every delivered register is then either consumed by exactly one later block of its holder, or \emph{stored} until the holder \emph{opens} it, which means measuring it in the computational basis. Every word to which we apply the analysis below is opened in one public round, its \emph{opening round}, in which every honest holder measures all its stored registers of the word; the outcomes are classical and become part of the history. The word is \emph{unopened} at a boundary if no honest holder has measured a register of it by then, and its opening \emph{ends} at the end of the opening round. We claim no secrecy for a word during its opening round, since the outcomes of honest measurements become public at once. Honest holders apply nothing else to the word's registers, and all encoders satisfy $D\ge p+3\sigma$.

We call these shares \emph{disposable}: once a register enters a completed block, its former holder retains no old input. A self-addressed output is a new share and is accounted for as such.

To \emph{reconstruct} the word, a requester sends one whole-word request to each current holder and collects its complete address--value reply under the request and quota rules of Section~3.2. Missing or malformed values at expected addresses become $0$. It then decodes the blocks in reverse order, treating each recovered input of a block as the readout of the output of the earlier block that produced it, and it ends with the input of the dealer block. Public addresses identify every block and position. No process other than the current holders is contacted; earlier blocks are decoded locally.

\paragraph{Marking.}
Marks at a boundary $s$ refer to the circuit as it stands at $s$. Its blocks are those whose round has begun by $s$, and a register that has been sent to a holder, or kept by it, and that is not an input of such a block counts as \emph{stored} at that holder, also before its delivery. At $s$, a block is \emph{exposed} if its holder is corrupted by $s$ and before the end of the block's round, and a stored register is exposed if its holder is corrupted by $s$. Marks are defined backward, starting from the stored registers. A stored register is \emph{marked} if it is exposed. A block is marked if it is exposed, or if more than $\sigma$ of its outputs are marked, where an output is marked if it is a marked stored register or an input of a marked block. The word is \emph{protected} at $s$ if its dealer block is unmarked. All these notions depend only on the history at $s$. They serve the analysis only; the protocol never identifies marked blocks.
The latent value $S_\omega$ labels the eventual outcomes; it is not a classical secret stored before opening. The proof constructs this label by showing that a uniform mixture of basis inputs gives the same relevant observations as the dealer's superposition. Protection at a fixed boundary is determined by the history, but continued protection depends on future corruptions. The following lemma therefore assumes protection on every continuation of $\hist_0$. We will satisfy this hypothesis by restricting the current committee to a budget-respecting continuation and using the already completed transfers below it.

\begin{lemma}[Disposable-Share Lemma]\label{lem:disposable}
Let $W$ be a set of words prepared by honest dealers and processed by honest holders through circuits of share encoders. Fix an adversary and a history $\hist_0$ at which every word of $W$ has been dealt and none has been opened. Suppose that, on every execution that extends $\hist_0$, every word of $W$ is protected at every boundary from $\hist_0$ up to the end of its last reconstruction. Then, on the executions that extend $\hist_0$, each $\omega\in W$ has a latent value $S_\omega\in\Fq$ such that:
\begin{enumerate}[label=(\alph*)]
\item for every history $\hist$ that extends $\hist_0$ and at which the words of $W'\subseteq W$ are not yet opened, \[\rho_{S_{W'}E|\hist}=q^{-|W'|}\textstyle\sum_{s}\ketbra ss\otimes\rho_{E|\hist};\]that is, the latent values are jointly uniform and independent of the history and of the adversary's quantum system;
\item on every execution that extends $\hist_0$, every requester that is honest through a reconstruction of $\omega$ obtains $S_\omega$.
\end{enumerate}
\end{lemma}

\begin{proof}\label{app:disposable}
The argument has two ingredients. First, tracing out the reliable outputs of protected blocks leaves the complementary systems independent of the dealer input. Second, for a basis input $s'$, the reliable readouts always decode to $s'$, whatever values are reported elsewhere. Consequently, distinct dealer values occupy disjoint decoding subspaces. This lets us replace the actual superposition by a uniform mixture for the histories, adversarial states and reconstruction outcomes under consideration, and use the mixture's basis label as the latent value.

We establish privacy and decoding before making this replacement. Privacy alone would not justify replacing a coherent input by a classical mixture. Basic quantum notation is collected in \cref{app:quantum}.

\paragraph{Tracking a history.} We use unnormalized states to track a history's probability and its conditional quantum state in the same calculation. Fix the adversary and the history $\hist_0$ of the lemma. For a history $\hist$, the operations applied along $\hist$, by honest processes and by the adversary, compose to a completely positive, trace-nonincreasing map $\Gamma_\hist$. Applied to the initial state, in which the dealer input of each word of $W$ is an operator $X_\omega$ and all other systems are in their fixed initial state, it gives the \emph{unnormalized branch state} of all systems at $\hist$. For the actual inputs $X_\omega=\ketbra{+_q}{+_q}$, its trace is $\Pr[\hist]$, and its partial trace over all systems held by uncorrupted processes is $\Pr[\hist]\,\rho_{E|\hist}$. All maps below are linear in the inputs $X_\omega$, and all identities between them hold with arbitrary reference systems attached.

\paragraph{Designated wires.} Fix a history $\hist$ that extends $\hist_0$, let $s$ be its boundary, fix a set $W'\subseteq W$ of words that are protected at $s$, and take all marks at $s$. For a word of $W'$, call a register a \emph{designated wire} while it is an input of an unmarked block or an unmarked stored register, from its delivery on, and while it is an unmarked output of an unmarked block, from the moment it is sent. Call the remaining systems the \emph{complementary subsystem}. Every system that the adversary holds at $s$ belongs to the complementary subsystem: a register of a word of $W'$ that a process corrupted by $s$ holds, or seized in transit, is an exposed stored register, an input of an exposed block or an output of an exposed block, and all of these are marked. Nothing acts on a designated wire except the unmarked block that consumes it. Its holder is not corrupted before the end of that block's round, or, for a stored register, by $s$; honest holders apply nothing else to it before the opening; and a register in transit can be touched only by corrupting its sender, whose block would then be exposed. 

This split groups systems only for the analysis. The complementary subsystem contains everything the adversary holds at $s$ and possibly more, for instance the inputs of a marked block whose holder is still honest. No register is duplicated: a marked block applied by an honest holder simply counts as an operation on the complementary subsystem, and the state of all systems is unchanged.

The branch map $\Gamma_\hist$ composes three kinds of operations in round order:
\begin{enumerate}[label=(\roman*)]
\item Completely positive maps on the complementary subsystem only, which include all adversarial operations, all marked blocks, all operations on other words and the measurements of opened words outside $W'$;
\item for each unmarked block $B$ of a word of $W'$, the isometry of its encoder applied to its input wires together with fresh $\ket{+_q}$ and $\ket0$ registers, whose outputs are designated wires except the at most $\sigma$ marked ones, which belong to the complementary subsystem;
\item relabellings, when an output of a marked block is delivered as an unmarked register and becomes a designated wire.
\end{enumerate}
If no word of $W'$ is opened at $\hist$, nothing else acts on designated wires; otherwise honest holders have also measured some of them.

\paragraph{Privacy by backward elimination.} Work backward through the unmarked blocks. Once a block's unmarked outputs are traced out, its at most $\sigma$ remaining outputs reveal nothing about its inputs, by the encoder privacy identity. Repeating this step removes the dealer input itself. The following induction makes this argument valid for each history, with arbitrary reference systems attached.

Suppose that no word of $W'$ is opened at $\hist$, and let $B_1,\dots,B_m$ be the unmarked blocks of the words of $W'$, ordered by round, with ties broken arbitrarily. For $0\le j\le m$, let $\Gamma^{(j)}_\hist$ be obtained from $\Gamma_\hist$ by replacing each of $B_{j+1},\dots,B_m$ with the map that traces out its input wires and prepares $q^{-|M_B|}I$ on the set $M_B$ of its marked outputs, and by deleting its unmarked outputs. Let $\Tr_{\mathrm{des}}$ trace out the designated wires at $s$, i.e., the unmarked stored registers. We show by downward induction on $j$ that
\begin{equation}\label{eq:app-invariant}
\Tr_{\mathrm{des}}\circ\Gamma_\hist=\Tr_{\mathrm{des}}\circ\Gamma^{(j)}_\hist .
\end{equation}
For $j=m$ there is nothing to show. For the step from $j$ to $j-1$, let $B=B_j$. Each unmarked output of $B$ is either an input of a later unmarked block, which lies in a later round and is therefore among $B_{j+1},\dots,B_m$, or an unmarked stored register. In $\Gamma^{(j)}_\hist$ the first kind is traced out by the replaced block in its round, and the second kind at the end. No map of kind (i) or (iii) acts on these wires, so the partial trace commutes with everything in between, and we may trace out all unmarked outputs of $B$ right after $B$. What remains of $B$ is the map from its input wires to $M_B$, which is $X\mapsto\Tr(X)\,q^{-|M_B|}I_{M_B}$ by \cref{lem:encoder}(a), since $|M_B|\le\sigma$. This is an identity of linear maps, so it holds with all other systems attached, and it is the replacement made in $\Gamma^{(j-1)}_\hist$. 

At $j=0$ every unmarked block is replaced, including the dealer blocks of the words of $W'$, which are unmarked because these words are protected. Their inputs are traced out, so for every joint input operator $X$ of these dealer blocks
\begin{equation}\label{eq:app-branch}
\Tr_{\mathrm{des}}\,\Gamma_\hist(X\otimes\cdots)=\Tr(X)\,\Omega_\hist,
\end{equation}
where the dots stand for the inputs of the other words and the initial state of all other systems, and $\Omega_\hist$ does not depend on $X$. Since the adversary's systems are among the complementary systems, the same holds for its state.

\paragraph{Decoding basis inputs.} Start with a basis value $s'$ at the dealer. An unmarked block can correct all of its marked outputs, so replacing its outputs by the recovered inputs preserves the recursively decoded dealer value. Operations on the complementary subsystem cannot change this conclusion, even if they are coherent. We express this property using the subspaces of readouts that decode to each value.

Now fix one word $\omega\in W'$, and let $\mathcal H$ be its unmarked stored registers at $s$. \emph{Recursive decoding} assigns a dealer value to values on the stored registers of $\omega$: it decodes the latest blocks first and treats each recovered input of a block as the value of the output that fed it, as described above. A string $y\in\Fq^{\mathcal H}$ is \emph{consistent with} $s'\in\Fq$ if recursive decoding returns $s'$ whatever the values on the stored registers outside $\mathcal H$ are. Distinct values have disjoint sets of consistent strings. Let $\Pi_{s'}$ be the projector, on the registers of $\mathcal H$, onto the span of the basis strings consistent with $s'$. We claim that for all $s_1,s_2\in\Fq$, with the dealer input $\ketbra{s_1}{s_2}$ for $\omega$ and any inputs for the other words,
\begin{equation}\label{eq:app-support}
\Gamma_\hist\big(\ketbra{s_1}{s_2}\otimes\cdots\big)=\Pi_{s_1}\,\Gamma_\hist\big(\ketbra{s_1}{s_2}\otimes\cdots\big)\,\Pi_{s_2}.
\end{equation}
Here we use the split above with $W'=\{\omega\}$; whether other words are opened does not matter, since only wires of $\omega$ are designated. Expand $\Gamma_\hist$ in Kraus operators along its composition, so that the left side is a sum of terms $K\ket{s_1}\ket{\phi}\bra{s_2}\bra{\phi'}K^\dagger$. It suffices to show that every vector $K\ket{s_1}\ket\phi$, expanded at any time in the computational basis of the designated wires that exist at that time, has only terms whose values decode recursively to $s_1$ whatever values the marked outputs take; at time $s$ these wires are exactly $\mathcal H$. 

We follow the composition. Before the dealer block, the only designated wire is the dealer input, with value $s_1$. A map of kind (i) acts only on the complementary subsystem, so it changes no basis value of a designated wire. A relabelling of kind (iii) adds a designated wire that is an output of a marked block; its value enters the decoding only through the decoding of that marked block, whose result is a marked output of an earlier block, so it does not affect the decoded dealer value. An unmarked block $B$ maps a basis term with input values $\mathbf u$ to a superposition over $\mathbf r$ of terms in which its outputs carry the codeword of $f_{\mathbf u,\mathbf r}$; at most $\sigma$ of them are marked, so by \cref{lem:encoder}(b) decoding $B$ returns $\mathbf u$ whatever the marked values are, and the decoded dealer value does not change. A measurement by an honest holder at the opening projects onto basis values and keeps the property. This proves \eqref{eq:app-support}. 

Amplitudes may cancel, but this only removes strings from the expansion; no step creates a string that decodes to a different value. The argument uses only how unmarked blocks act on computational-basis values and that nothing else acts on designated wires; the adversary's operations may be arbitrary and coherent.

\paragraph{From the superposition to latent values.} The decoding subspaces for distinct values are orthogonal. Before opening, tracing out their registers removes cross terms from the adversary's state; at opening, the measured readouts remove those terms from the full branch state. We now use these two facts to justify the latent-value interpretation.

We compare the dealer input $\ket{+_q}$ of $\omega$ with the uniform mixture $q^{-1}\sum_{s'}\ketbra{s'}{s'}$; they differ by the cross terms $q^{-1}\ketbra{s_1}{s_2}$ with $s_1\ne s_2$. Let $\hist$ extend $\hist_0$. If $\hist$ ends no later than the end of the opening of $\omega$, then $\omega$ is protected at $\hist$ by hypothesis, and by \eqref{eq:app-support} the contribution of a cross term to the adversarial state is
\[
\Tr_{\mathrm{des}}\big(\Pi_{s_1}\Gamma_\hist(\cdots)\Pi_{s_2}\big)=\Tr_{\mathrm{des}}\big(\Pi_{s_2}\Pi_{s_1}\Gamma_\hist(\cdots)\big)=0,
\]
because $\Pi_{s_1}$ and $\Pi_{s_2}$ act on traced registers and $\Pi_{s_2}\Pi_{s_1}=0$. At the end of the opening, every holder that is honest at that boundary has measured all its stored registers of $\omega$, and these registers are all of $\mathcal H$. So the branch state of a history at that boundary has the form $P_y\Gamma(\cdots)P_y$ for the measured string $y$ on $\mathcal H$, and by \eqref{eq:app-support} the cross term vanishes on all systems, since $y$ is not consistent with both $s_1$ and $s_2$. Every later history extends such a history, so cross terms vanish there as well, whether or not $\omega$ stays protected later. 

Hence, on the executions that extend $\hist_0$, the histories and the adversary's states have the same joint distribution for the input $\ket{+_q}$ and for a uniformly random basis input; in the second case, let $S_\omega$ be the basis input. 

For several words, a cross term of the joint input is off-diagonal in some coordinate $\omega$, and the same argument applies to that coordinate with arbitrary operators on the others, so we may replace all inputs at once. Finally, if no word of $W'$ is opened at $\hist$, then \eqref{eq:app-branch} with $X=\bigotimes_{\omega\in W'}\ketbra{s_\omega}{s_\omega}$ shows that the adversary's unnormalized state on the branch of $\hist$ together with $S_{W'}=(s_\omega)_\omega$ is $q^{-|W'|}$ times a fixed operator, for every value $(s_\omega)_\omega$. This is part (a), and it covers words of $W\setminus W'$ that are already opened at $\hist$, since their circuits belong to the complementary subsystem in this partition.

\paragraph{Reconstruction.} Take the basis inputs of the previous paragraph, and consider an execution that extends $\hist_0$ and a requester that is honest through a reconstruction of $\omega$. Let $s$ be the end of the reconstruction; by hypothesis $\omega$ is protected at $s$, and let $\mathcal H$ be its unmarked stored registers at $s$. Their holders are honest through the reconstruction and report the values they measured at the opening; by \eqref{eq:app-support} these form a string consistent with $S_\omega$. Every other report concerns a stored register outside $\mathcal H$, whose holder is corrupted by $s$, or is missing and replaced by $0$. Since the string on $\mathcal H$ is consistent with $S_\omega$, recursive decoding returns $S_\omega$ whatever these other values are. 
\end{proof}

\subsection{Protection along election paths}

Like the committees, the forwarding graphs are fixed in advance by exhaustive search and known to all processes.

\begin{lemma}[Forwarding graphs; proof in \cref{app:uplinks}]\label{lem:uplinks}
Let $\rho\in(0,\frac12]$, and let $U,V$ be committees of sizes $k\le k'$. We call the vertices of $U$ \emph{rows} and those of $V$ \emph{columns}. Let $D\le k'$ be an integer with $D=k'$ or $2D\rho^3k\ge(k+k')\ln2+\ln8$. There is a fixed public bipartite graph between $U$ and $V$ with the following properties:
\begin{enumerate}[label=(\roman*)]
\item every row has $D$ distinct neighbors;
\item every column has degree at most $J_{\mathrm{graph}}:=\lceil\max\{2ekD/k',3L\}\rceil$ (or $J_{\mathrm{graph}}=k$ if $D=k'$);
\item for every $Z\subseteq V$,
\begin{equation}\label{eq:sampler}
\#\Big\{u\in U:\ \frac{|N(u)\cap Z|}{D}>\frac{|Z|}{k'}+\rho\Big\}\le\rho k .
\end{equation}
\end{enumerate}
\end{lemma}

Consider an array whose registers have moved along committees $C_1,\dots,C_\ell$, where $C_1$ received the dealer's shares and $C_{j+1}$ is the parent committee of $C_j$. For $u\in C_j$, let $N(u)\subseteq C_{j+1}$ be its set of forwarding neighbors, and let $D_j=|N(u)|$ be the number of forwarding edges of each member of $C_j$.

\begin{lemma}[Protection along a path]\label{lem:path}
Consider a good array whose registers have moved along committees $C_1,\dots,C_\ell$, with forwarding graphs satisfying \eqref{eq:sampler} for sampler error $\rho$, and with $\gamma+2\rho+D_j^{-1}<\theta$ for all $j$ and $(\gamma+\rho)|C_1|\le\sigma(|C_1|)$. Let $s$ be a boundary no earlier than the end of the round in which the array was transferred out of $C_{\ell-1}$ (the distribution round if $\ell=1$). Make two assumptions:
\begin{itemize}
\item for each $j<\ell$, at most $\gamma|C_j|$ members of $C_j$ were corrupted by the end of the round in which the array was transferred out of $C_j$, and all members of $C_j$ honest at the end of that round transferred it;
\item at most $\gamma|C_\ell|$ members of $C_\ell$ are corrupted by $s$.
\end{itemize}
Then every word of the array that reached $C_\ell$, and whose transfer out of $C_\ell$ has not begun by $s$, is protected at $s$, whether or not it has been opened.
\end{lemma}

\begin{proof}
For $j\le\ell$, let $Z_j$ be the fault set in $C_j$ at the boundary named in the assumptions, i.e., at the end of the transfer round out of $C_j$ if $j<\ell$, and at $s$ if $j=\ell$. All marks are taken at $s$. We bound possible marked blocks by sets $Y_j$: these include corrupted holders and holders with too many outputs leading into the next such set. Starting from $Y_\ell=Z_\ell$, define backward
\[
Y_j=Z_j\cup\{u\in C_j:\ |N(u)\cap Y_{j+1}|>\sigma(D_j)\}\qquad\text{for }j<\ell .
\]

We prove together, by backward induction, that $|Y_j|\le(\gamma+\rho)|C_j|$ and that every block outside $Y_j$ is unmarked. At the final committee, $Y_\ell=Z_\ell$ has size at most $\gamma|C_\ell|$. A holder outside it is not corrupted by $s$, so its stored registers are unmarked, whether or not they have been measured.

For the induction step, let $j<\ell$ and $u\in C_j\setminus Y_j$. Since $u\notin Z_j$, it was not corrupted before the end of its transfer round, so none of its blocks is exposed. Each of its blocks sends one output per forwarding edge (line~\ref{ln:dt-send}), and $u$ has at most $\sigma(D_j)$ forwarding edges into $Y_{j+1}$, so at most $\sigma(D_j)$ outputs of each block reach $Y_{j+1}$. The outputs that reach holders outside $Y_{j+1}$ are unmarked by induction; if $j+1<\ell$, such a holder transferred the array, so the output is an input of one of its blocks. Hence no block of $u$ is marked.

For the size bound in the same induction step, apply \eqref{eq:sampler} to the forwarding graph from $C_j$ to $C_{j+1}$ with $Z=Y_{j+1}$. All but at most $\rho|C_j|$ rows $u$ satisfy
\[
|N(u)\cap Y_{j+1}|\ \le\ (\gamma+2\rho)D_j\ <\ \theta D_j-1\ <\ \sigma(D_j),
\]
by the assumption $\gamma+2\rho+D_j^{-1}<\theta$, so they enter $Y_j$ only if they lie in $Z_j$. Adding $|Z_j|\le\gamma|C_j|$ proves the bound.

Finally, the dealer block of each such word is not exposed, since the array is good. It sends one output to each member of $C_1$, so at most $|Y_1|\le(\gamma+\rho)|C_1|\le\sigma(|C_1|)$ of its outputs reach $Y_1$, and its other outputs are unmarked. Hence the dealer block is unmarked, i.e., the word is protected at $s$.
\end{proof}

Only faults present by a completed transfer enter its fault set. Later corruption cannot reveal the inputs consumed by that transfer. This is the reason the path argument remains valid under adaptive corruption.

\begin{lemma}[The sharing scheme of \QBAP{} provides fresh common coins]\label{lem:moving-coins}
Suppose that every forwarding graph has properties (i)--(iii) of \cref{lem:uplinks} for sampler error $\rho$, that $\gamma+2\rho+D_\ell^{-1}<\theta$ and $p_\ell+3\sigma(D_\ell)\le D_\ell$ for every $\ell$, and that $(\gamma+\rho)k_1\le\sigma(k_1)$ unless the level-$1$ committees are $P$. Then the sharing scheme of \QBAP{} provides fresh common coins in the sense of \cref{def:store}, up to the first honest decision in epochs entered before it, with the committee budgets specified above.
\end{lemma}

\begin{proof}[Proof of \cref{lem:moving-coins}]
Fix an epoch-entry history preceding the first honest decision. Every uncorrupted process is still active, so every good array is actually distributed by an honest dealer. Every encoder used in distribution or \cref{alg:transfer} satisfies $D\ge p+3\sigma$: blocks with $p=1$ do, since $3\sigma(D)=3\lfloor\theta D\rfloor<D$, and packed blocks by assumption. So every word is processed through a circuit of share encoders as in \cref{lem:disposable}. 

Successful ancestors guarantee completed transfers, the current committee budget guarantees continued protection, and the Disposable-Share Lemma then supplies the two interface properties. First consider the case in which the level-$1$ committees are not $P$.

Fix a level-$\ell$ node $v$, an adversary under which $\Bud_v$ holds on every execution, and a history $\hist_0$ at the common deadline of level $\ell-1$; it determines the set $W$ of \cref{def:store}, whose words are unopened at $\hist_0$. Let $x$ be an array of $G_{\ell-1}$ in $v$'s subtree; it is good, and the nodes $u_1,\dots,u_{\ell-1}$ below $v$ on its path are successful and have $x$ in their common lists. So, at the delivery boundary $d_{u_j}$, at most $\gamma|C_{u_j}|$ members of $C_{u_j}$ are corrupted, and every member honest then has $x$ in its own output list and transferred it (line~\ref{ln:dt-own}). These facts are fixed by $\hist_0$.

The words of $W$ are opened and reconstructed by members of $C_v$ before $d_v$, since holders ignore wrong-stage requests, and they are never transferred out of $C_v$; and $\Bud_v$ bounds the corruption of $C_v$ at every boundary up to $d_v$. Hence \cref{lem:path}, applied at every boundary from $\hist_0$ to $d_v$, protects every word of $W$; if the path reaches $P$ below $v$, we apply it to the path up to the first committee equal to $P$, which never has more than $t\le\gamma n$ corrupted members. \Cref{lem:disposable} then gives (a) and (b) for $W$ and $\hist_0$. For the final words, the root committee is $P$, whose corruption budget always holds. To apply \cref{lem:disposable} without assuming that actual holders remain active after deciding, use an analysis-only extension: after $s_{\mathrm{dec}}$, keep uncorrupted holders' registers and let them complete all scheduled final-word openings and replies for this epoch, despite their actual halting rule. Extend the adversary arbitrarily subject to the global corruption bound. This extension is identical to the actual protocol strictly before $s_{\mathrm{dec}}$, and every counted final word remains protected throughout its reconstruction in the extension. The Disposable-Share Lemma supplies joint latent values there; restricting its conclusions to histories and reconstructions before $s_{\mathrm{dec}}$ gives exactly \cref{def:store}. This is an extension of the stopped execution, not conditioning on a future decision or on future corruption choices. No freshness claim is made for actual post-decision reconstructions.

If the level-$1$ committees are $P$, the circuit of each word is a single dealer block $V_{n,\sigma(n),1}$ with $\sigma(n)\ge t$, the registers never move (\cref{sec:transfer-conventions}), and the argument of \cref{cor:single-block} applies with $\sigma(n)$ in place of $t$, using the same analysis-only extension for final words.
\end{proof}

\section{Parameters and complexity}\label{sec:proof-packed}

\subsection{Requirements}
The choices must satisfy three sets of conditions. First, committee assignments must meet \cref{lem:committees}, with the root committee equal to $P$. Second, every forwarding graph must satisfy \cref{lem:uplinks}, have in-degree at most $J$, and meet the protection and decoding inequalities
\[
\gamma+2\rho+D_\ell^{-1}<\theta,\qquad
p_\ell+3\sigma(D_\ell)\le D_\ell,\qquad
(\gamma+\rho)k_1\le\sigma(k_1).
\]
The first two apply at levels with transfers; the last is needed only when the initial committee is not $P$. If it is $P$, the single-block argument applies with $\sigma(n)\ge t$.

Third, the losses and list length must leave a good majority at the root and a small epoch failure probability:
\[
F_0-h(\beta+\tau+\eta)>\tfrac12,\qquad
p_n=\Big(u2^{-\tau r}+be^{-\eta^2w/2}\Big)\sum_{\ell=1}^{h}m_\ell+2^{-w/2}\le n^{-c}.
\]
Here $c>0$ sets the desired failure exponent. The headline algorithm uses $c=1$. For $h=0$ there are no elections or transfers, and the sum is zero.

\subsection{Parameter choices}

For $n=1$, the single process decides its input at once. For $n\ge2$, we have $L\ge1$ and use the following choices for the failure exponent $c>0$:
\begin{gather}
b=2^{\lceil\log_2(10/\varepsilon)\rceil},\qquad H=\lceil\log_b(10n)\rceil,\qquad \beta=\eta=\frac1{20H},\qquad \tau=\frac1{100H},\qquad \rho=\frac\varepsilon{10},\label{eq:params-packed}\\
A=10^6(c+2)\,\varepsilon^{-2},\qquad w=2^{\lceil\log_2(AH^2L)\rceil},\qquad h=\max\{0,\lceil\log_b(10n/w)\rceil\},\qquad N=wb^h,\notag\\
k_\ell=\begin{cases}\kappa b^{\ell-1}&\text{if }\kappa b^{\ell-1}\le n/b,\\ n&\text{otherwise,}\end{cases}\qquad \kappa=\lceil K(m_1)\rceil,\notag\\
C_\varepsilon=\lceil2/\rho^3\rceil,\qquad D_\ell=\lceil C_\varepsilon L\,k_{\ell+1}/k_\ell\rceil,\qquad J=\lceil6C_\varepsilon L\rceil .\notag
\end{gather}
The quantities $m_1$, $\kappa$, and the election committee sizes are used only when $h\ge1$; when $h=0$, the direct distribution described below applies. Here $D_\ell$ is defined only when $1\le\ell<h$ and $k_\ell<n$, and \cref{app:checks-packed} shows that these forwarding graphs are never complete and have in-degree at most $J$. The base size $\kappa$ need not be the size of the level-$1$ committees: $k_1=\kappa$ if $\kappa\le n/b$, and $k_1=n$ otherwise. Committee sizes grow exactly by the factor $b$ per level until they exceed $n/b$; from then on every committee is $P$, and shares stop moving (\cref{sec:transfer-conventions}). In particular $k_h=n$ (\cref{app:checks-packed}). If $h=0$, there are no committees below the root, we set $k_1=n$, and every word is dealt to $P$.

\subsection{Verification of the requirements}
Appendix~\ref{app:checks-packed} checks the requirements in the same order: committee sizes and the root, forwarding graphs and block capacity, the root's good fraction, and the failure probability. In particular, it gives $F_0-h(\beta+\tau+\eta)>0.52$ and $p_n\le n^{-c}$. The checks include rounding, the first capped level, and the case $h=0$. Thus \cref{lem:moving-coins} supplies the randomness interface, and \cref{cor:generic} applies. It remains to bound the communication of an epoch and combine that bound with its progress probability.

\subsection{Cost accounting}\label{app:costs}

For $h\ge1$, we count the qubits, classical bits and rounds of one epoch of \QBAP{}, with the parameters \eqref{eq:params-packed}. All bounds below hold deterministically for one epoch, also in failed elections, because they rely only on the per-sender, per-slot and per-requester caps. Throughout, $d,\iota=O(L)$, and requests and replies carry $O(L)$-bit headers; each returned register additionally carries an $O(L)$-bit public address. Recall that level $\ell$ has $m_\ell$ nodes with committees of size $k_\ell$. Thus $m_\ell k_\ell$ is the number of memberships at level $\ell$, and $m_\ell k_\ell^2$ bounds the number of messages of one all-to-all exchange inside the level's committees.

\paragraph{Committee sums.} Two sums control the accounting: $\sum_\ell m_\ell k_\ell$ counts holder memberships, and $\sum_\ell m_\ell k_\ell^2$ bounds the communicating pairs in committee exchanges. Below the cap, the first quantity per level is constant and committee sizes grow geometrically; above the cap, the number of committees decreases geometrically. We use these facts to bound both sums.

A level is \emph{capped} if $k_\ell=n$. If some level is uncapped, then $\kappa\le n/b$, so $\beta m_1\ge1$, since otherwise $K(m_1)>n$, and
\[
m_1\kappa\le\frac2{\varepsilon^2}\Big(\frac{n\ln2+2}\beta+m_1\ln\frac e\beta\Big)+m_1=O_\varepsilon(nH).
\]
Every uncapped level has $m_\ell k_\ell=m_1\kappa$, because $m_\ell=m_1/b^{\ell-1}$ and $k_\ell=\kappa b^{\ell-1}$. The uncapped sizes grow by the factor $b\ge2$ and stay at most $n/b$, so $\sum_{\text{uncapped}}k_\ell\le n$. At the first capped level, $\kappa b^{\ell-1}>n/b$ gives $m_\ell<m_1\kappa b/n=O_\varepsilon(H)$, and $m_\ell$ decreases by the factor $b$ per level from there, so the capped levels have $O_\varepsilon(H)$ nodes in total. If $\kappa>n/b$, all levels are capped, and $K(m_1)>n/b-1$ forces $m_1=O_\varepsilon(H)$ for $n$ larger than a constant depending on $\varepsilon$, with the same conclusion. Hence
\[
\textstyle\sum_\ell m_\ell k_\ell\le h\,m_1\kappa+O_\varepsilon(H)\,n=O_\varepsilon(nH^2),\quad\sum_\ell m_\ell k_\ell^2\le m_1\kappa\sum_{\text{uncapped}}k_\ell+O_\varepsilon(H)\,n^2=O_\varepsilon(n^2H).
\]

\paragraph{One epoch.} A holder packs its registers of a word into a single block before each transfer, so at every level it has at most $J$ registers of a word. We have $b=\Theta(1/\varepsilon)$, $H=O(L)$, $w=O_{\varepsilon,c}(L^3)$, $M=2h+1$, and $D_\ell\le C_\varepsilon Lb^2+1=O_\varepsilon(L)$, since every transfer has $k'/k\le b^2$. Also $J=\lceil6C_\varepsilon L\rceil=O_\varepsilon(L)$.
\begin{itemize}
\item \emph{Qubits:} the initial sharing costs $NMk_1d=m_1k_1\cdot bw\cdot Md=O_{\varepsilon,c}(nL^6)$, since $m_1k_1=O_\varepsilon(nH)$ whether or not level $1$ is capped. Per transfer level, $m_\ell k_\ell\cdot wM\cdot D_\ell d=O_{\varepsilon,c}(nL\cdot L^3\cdot L\cdot L\cdot L)$. Over $h$ levels this is $O_{\varepsilon,c}(nL^8)$.
\item \emph{Reconstruction:} at most $2r$ words per membership, each using one request per holder and returning at most $k_\ell J$ address--value pairs. Each pair has $O(L)$ bits, including its field element and address, so the cost is $O(rJL\sum_\ell m_\ell k_\ell^2)=O_{\varepsilon,c}(n^2L^6)$ bits. Root reconstruction, including addresses, costs $O(n\cdot w\cdot nJL)=O_{\varepsilon,c}(n^2L^5)$.
\item \emph{Votes:} each communicating pair exchanges $u$-bit vectors during $r$ phases, giving $O(\sum_\ell m_\ell k_\ell^2\cdot ru)=O_{\varepsilon,c}(n^2L^8)$ bits.
\item \emph{Lists:} each membership sends a $w$-slot list of $O(wL)$ bits to at most $n$ parent members, giving $O(\sum_\ell m_\ell k_\ell\cdot n\cdot wL)=O_{\varepsilon,c}(n^2L^6)$ bits.
\item \emph{\FinalAg{}:} $O(n^2wL)$ bits.
\item \emph{Rounds:} packing adds local gates but no round, so $1+h(4bw+3)+4(w+1)=O_{\varepsilon,c}(L^4)$.
\end{itemize}
If $h=0$, distribution uses $O(nw\log n)$ qubits and final agreement uses $O(n^2w\log n)$ bits in $O(w)$ rounds, also within the stated bounds. For $h\ge1$, agreement votes dominate the classical communication. Thus an epoch uses $O_{\varepsilon,c}(nL^8)$ qubits, $O_{\varepsilon,c}(n^2L^8)$ classical bits and $O_{\varepsilon,c}(L^4)$ rounds.

\subsection{Completion of the upper bound}
\begin{proof}[Proof of \cref{thm:packed}]
If $n=1$, the process decides its input immediately and sends nothing. For $n\ge2$, use the choices above with $c=1$. The verified committee and protection conditions establish the randomness interface, and the root inequality allows us to apply \cref{cor:generic}. Agreement and validity hold on every execution. Conditional on any epoch-entry history before the first honest decision, the failure probability is at most $p_n\le1/n\le1/2$. Hence termination holds with probability~$1$, and the expected number of epochs through the last correct halt is at most $(1-p_n)^{-1}+2\le4$.

The deterministic per-epoch bounds of \cref{app:costs} hold even when elections fail. Multiplying them by the expected number of epochs gives $O_\varepsilon(n\log^8 n)$ qubits, $O_\varepsilon(n^2\log^8 n)$ classical bits, and $O_\varepsilon(\log^4 n)$ rounds.
\end{proof}
\section{A quadratic lower bound}\label{sec:lower}

We prove the lower bound by adapting the message-suppression argument of Abraham et al.~\cite{AbrahamCDNPRS19}. The quantum step is to compare suppressing a register with delivering it to a process that discards it. Both operations trace out that register and therefore have the same effect on the remaining systems.

A \emph{transmission} is a channel use, classical or quantum. By the model, every channel use carries at least one bit or one qubit and is charged for it, so every transmission sent by an honest process contributes at least $1$ to $C+Q$.

\begin{proof}
If communication is too small, some process can be isolated within the corruption budget. Its output distribution is the same for two different inputs of the other processes, yet correctness will require that distribution to give each bit probability greater than one half. The case $K=\infty$ is immediate, so assume $K<\infty$.

Reserve $a-1$ corruptions for initially faulty processes and $s$ for suppressing transmissions to one additional process. Balancing these two quantities will maximize the bound. Choose positive integers $a$ and $s$ with $a+s=t+1$ and $as=\lfloor(t+1)^2/4\rfloor$. Fix a set $V$ of $a$ processes, let $U=P\setminus V$, and fix an input $x_v$ for each $v\in V$.

\emph{Reference executions.} For $b\in\{0,1\}$, let $E_b$ be the execution in which every process of $U$ has input $b$, and the adversary corrupts the processes of $V$ at the start. Each $v\in V$ then runs the algorithm with input $x_v$, except that it ignores every transmission from $U$: it discards the contents and behaves as if the transmission had not arrived, so neither the payloads nor the arrival of transmissions from $U$ affect it. The adversary does nothing else. The correct processes are those of $U$, and they all have input $b$, so by validity and termination every process of $U$ decides $b$ with probability~$1$.

Let $N_{b,v}$ count transmissions from $U$ to $v$ in $E_b$. The senders are correct and send nothing after the last correct halt, so all these transmissions contribute to $C+Q$. Thus $\sum_{v\in V}\E[N_{b,v}]\le K$ for each $b$. Averaging over $V$, choose $p$ with $\E[N_{0,p}]+\E[N_{1,p}]\le2K/a$. Suppose, for contradiction, that $K<as/4$. Each expectation is less than $s/2$, and Markov's inequality gives, for $H_b=\{N_{b,p}\le s\}$,
\begin{equation}\label{eq:lb-markov}
\Pr[H_b]\ \ge\ 1-\frac{\E[N_{b,p}]}{s+1}\ >\ \frac12 .
\end{equation}

No process of $V$ is affected by transmissions from $U$, neither by their contents nor by their arrival, and the processes share no initial entanglement. The processes of $V$ may send registers to $U$, possibly entangled with their own, but nothing that depends on these registers comes back to $V$, and operations on systems held by $U$ cannot change the state of $V$. Hence, by no-signalling, the joint evolution of the processes of $V$ in $E_b$ does not depend on $b$. In particular, the output $O_p\in\{0,1,\bot\}$ that $p$ computes in $E_b$, with $\bot$ if it does not decide, has a distribution $\mu$ that is the same for $b=0$ and $b=1$.

\emph{Isolating $p$.} Now let $E'_b$ be the execution with the same inputs, in which $p$ has input $x_p$ and is never corrupted. The adversary corrupts $V\setminus\{p\}$ at the start and runs these processes exactly as in $E_b$. Whenever a process of $U$ attempts one of the first $s$ transmissions from $U$ to $p$, the adversary sees the attempt, corrupts the sender in that round, seizes the transmission and discards it. A corrupted sender otherwise keeps running the algorithm. At most $(a-1)+s=t$ processes are corrupted, and at least $n-t-1\ge1$ processes of $U$ stay correct.

We claim that $E'_b$ and $E_b$ have the same distribution until a possible $(s+1)$st transmission from $U$ to $p$. The processes of $V\setminus\{p\}$ act in the same way in both, and $p$ runs the algorithm with input $x_p$ and nothing from $U$ in both. 

It remains to compare a delivered and a seized transmission. Let the sender apply a channel $\Phi$ whose output is its new state $S'$ and the outgoing system $M$. In $E_b$, $M$ is delivered to $p$, which discards it; in $E'_b$, the adversary seizes $M$ and discards it. In both cases the remaining systems undergo the channel $\Tr_M\circ\Phi$, also when the sender is entangled with other systems, and the classical information that the adversary observes is the same. This proves the claim.

On $H_b$ there is no $(s+1)$st transmission, so on $H_b$ the two executions coincide. Thus the processes of $U$ that stay correct in $E'_b$ decide $b$, and $p$ outputs $O_p$. Since $p$ is correct in $E'_b$, agreement and termination give $O_p=b$ with probability~$1$ on $H_b$. With \eqref{eq:lb-markov}, $\mu(b)=\Pr[O_p=b]\ge\Pr[H_b]>1/2$ for both $b\in\{0,1\}$, which contradicts $\mu(0)+\mu(1)\le1$. Hence $K\ge as/4=\frac14\lfloor(t+1)^2/4\rfloor$.
\end{proof}

The proof uses only two properties of the model: the adversary may suppress a message in transit by corrupting its sender during the round, and it sees which channels are used. The first one is essential already in the classical argument~\cite{AbrahamCDNPRS19}, and the subquadratic protocol of King and Saia~\cite{KingSaia11} works against an adversary that lacks it. Our algorithms are analyzed against this stronger adversary, so \cref{thm:lower} applies to them, and \cref{cor:optimal} follows.

\begin{proof}[Proof of \cref{cor:optimal}]
For $t=\Theta(n)$, \cref{thm:lower} gives $K=\Omega(n^2)$. Since $\E[C+Q]=\E[C]+\E[Q]$ for every input vector and adversary, $K\le\sup_{\mathbf x,\mathcal A}\E[C]+\sup_{\mathbf x,\mathcal A}\E[Q]$, so $\sup_{\mathbf x,\mathcal A}\E[C]\ge K-\sup_{\mathbf x,\mathcal A}\E[Q]=\Omega(n^2)$ when $\sup_{\mathbf x,\mathcal A}\E[Q]=o(n^2)$. 

For \QBAP{}, the bounds of \cref{thm:packed} hold uniformly over input vectors and adversaries, so $\sup_{\mathbf x,\mathcal A}\E[Q+C]=O(n^2\log^8n)$, $\sup_{\mathbf x,\mathcal A}\E[C]=O(n^2\log^8n)$ and $\sup_{\mathbf x,\mathcal A}\E[Q]=O(n\log^8n)=o(n^2)$.
\end{proof}
\section{Conclusions and open problems}\label{sec:conclusions}

The construction separates the selection of hidden randomness from its quantum representation. The election analysis applies to any sharing scheme satisfying the freshness interface, while the Disposable-Share Lemma gives a way to implement it under adaptive corruption. Several questions remain.

Ben-Or and Hassidim~\cite{BenOrHassidim05} reach agreement in $O(1)$ expected rounds, while \QBAP{} needs $O(\log^4n)$. Our round count comes from running one coin phase per candidate slot of an election, which gives $\Theta(\log^3n)$ phases per level over $\Theta(\log n)$ levels. Is there a protocol with $O(1)$, or at least substantially fewer, expected rounds that keeps the total communication at $\Ot(n^2)$ and the quantum communication at $\Ot(n)$?

\Cref{thm:lower} bounds the total communication, but not the qubits alone. Without quantum communication, classical full-information protocols need $\Omega(t^2/(n\log n))$ expected rounds~\cite{Efron26}. Is $\Omega(n)$ quantum communication necessary for polylogarithmic rounds, or more generally, what is the trade-off between qubits and rounds?

Our analysis requires $t\le(1/3-\varepsilon)n$, and the constants in our bounds are explicit but large. Reaching $t<n/3$ and reducing these constants are natural goals.

\section*{Acknowledgments}
\addcontentsline{toc}{section}{Acknowledgments}
The author thanks Dariusz Kowalski for introducing him to the problem of quantum Byzantine agreement and for countless hours of discussion that deepened his understanding of the problem.

\appendix
\section{Quantum notation and algebra}\label{app:proofs-prelim}\label{app:quantum}
\paragraph{Quantum notation.} A density operator is positive semidefinite with trace one. The adversary's operations, conditioned on a classical outcome, form a completely positive, trace-nonincreasing linear map, and the trace of its output is the probability of that outcome. A quantum instrument is a family of such maps whose sum preserves trace. Composing the operations along a fixed history gives again such a map. For a joint system on $A$ and $\mathcal{H}$ we have
\begin{equation}\label{eq:ptrace}
\Tr_{\mathcal{H}}\big(\ketbra{x}{y}\otimes\ketbra{u}{v}\big)=\langle v|u\rangle\,\ketbra{x}{y},
\end{equation}
so orthogonal states of a discarded subsystem remove the corresponding off-diagonal terms.

We also recall what independence from a quantum adversary means. A classical--quantum state $\rho_{SE}=\sum_sp_s\ketbra{s}{s}\otimes\rho^s_E$ satisfies $\rho_{SE}=\rho_S\otimes\rho_E$ exactly when $S$ is independent of the adversary's whole quantum system, and hence of any measurement that the adversary later performs on it. This is stronger than independence from a particular classical transcript.

\subsection{Algebra}\label{app:algebra}

We use the standard distance and unique-decoding properties of Reed--Solomon codes~\cite{ReedSolomon60,McElieceSarwate81}.

\begin{lemma}[Roots and distance]\label{lem:distance}
Two distinct polynomials of degree at most $t$ agree on at most $t$ points. Hence the evaluation code of degree-$t$ polynomials on $D$ distinct points has minimum distance $D-t$. If $D\ge t+1+2z$, a vector within Hamming distance $z$ of a codeword determines that codeword uniquely.
\end{lemma}

\begin{proof}
The difference of two distinct degree-at-most-$t$ polynomials is nonzero and has at most $t$ roots, so their evaluation vectors differ in at least $D-t$ positions. If a received vector were within distance $z$ of two codewords, those codewords would be at distance at most $2z$, contradicting $D-t\ge2z+1$.
\end{proof}

\section{Single-block sharing}\label{sec:global}

The global-sharing example uses a single encoder block. An honest dealer applies $V_{n,t,1}$ to $\ket{+_q}$ and sends output $i$ to process $i$, retaining only its own output share. Holders store their shares until the scheduled opening, then measure them in the computational basis. A requester sends requests in one round and receives readouts in the next. It replaces missing reports by zero, decodes the degree-at-most-$t$ polynomial, and returns its constant coefficient, or zero if decoding fails. No old dealer input remains. This is the single-block instance of the sharing construction~\cite{CleveGottesmanLo99,Gottesman00}.

\begin{corollary}[Single-block sharing]\label[corollary]{cor:single-block}
Consider words distributed by the single-block construction above.
\begin{enumerate}[label=(\alph*)]
\item Let $\hist_0$ be a history, at the end of the distribution round or later, at which none of these words has been opened and at which their dealers were uncorrupted at the end of the distribution round. Then, for every adversary, the conclusions of \cref{lem:disposable} hold for these words and $\hist_0$.
\item In any case, a reconstruction returns a field element within two rounds; if a dealer was corrupted earlier, the values of its word need not be common or unbiased.
\item Preparing a word costs $O(n\log n)$ qubits, and reconstructing it to one requester costs $O(n\log n)$ classical bits.
\end{enumerate}
\end{corollary}

\begin{proof}
The dealer block remains unmarked because it is honest and at most $t$ of its outputs can ever be exposed. More precisely, it satisfies $n\ge1+3t$ and is not exposed because the dealer stays honest through distribution. At most $t$ output shares can ever be exposed, so the block remains unmarked on every continuation. The Disposable-Share Lemma gives (a). The same argument applies to a single block $V_{n,\sigma(n),1}$ when $t\le\sigma(n)$ and $1+3\sigma(n)\le n$.

Requests and replies take two rounds; missing reports are replaced by zero, and a failed decoding returns zero. This proves (b) even for faulty dealers. Distribution sends $n-1$ registers of $O(\log n)$ qubits. Each requester sends $n$ requests and receives at most $n$ field-element reports, with $O(\log n)$-bit headers, giving (c).
\end{proof}

\section{Concentration bounds, committees and forwarding graphs}\label{app:conc}

This appendix proves the existence of the public committees of \cref{lem:committees} and of the forwarding graphs of \cref{lem:uplinks}. Both proofs pick a random object, show that it has the required properties with positive probability, and then fix a good object by exhaustive search. We first collect the concentration bounds they use.

\subsection{Concentration}

We use three standard bounds. The first is Hoeffding's inequality for sampling without replacement~\cite[Theorems~2 and~4]{Hoeffding63}, the second is the standard binomial upper bound, and the third is the multiplicative Chernoff bound~\cite[Chapter~4]{MitzenmacherUpfal17}.

\begin{lemma}[Sampling without replacement]\label{lem:wor}
Let a population of $n$ binary values contain a fraction $p$ of ones, and let $X$ count the ones in a uniform sample of $k$ distinct positions. Then $\Pr[X-kp\ge y]\le\exp(-2y^2/k)$ for $y\ge0$.
\end{lemma}

\begin{lemma}[Binomial upper bound]\label{lem:binom-upper}
If $X$ is a sum of $m$ independent indicators, each with probability $p$, and $\mu=mp$, then $\Pr[X\ge d]\le\binom md p^d\le(e\mu/d)^d$ for every positive integer $d\le m$. In particular, $\Pr[X\ge d]\le2^{-d}$ for $d\ge2e\mu$.
\end{lemma}

\begin{lemma}[Binomial lower tail]\label{lem:chernoff}
Let $X$ be a sum of independent indicators with mean $\mu>0$. For $0\le\zeta\le1$, $\Pr[X<(1-\zeta)\mu]\le e^{-\zeta^2\mu/2}$.
\end{lemma}

\subsection{Proof of \texorpdfstring{\cref{lem:committees}}{the committee lemma}}\label{app:committees}

Recall that the committee budget uses $\gamma=1/3-\varepsilon/2$. If $k=n$, assign $P$ to every node. No committee exceeds its budget because $t\le\gamma n$, and every process lies in all $m\le\Delta$ committees.

Otherwise $k<n$. Choose $m$ independent uniform $k$-subsets of $P$, one per node; this sampling is used only for the existence proof. For a fixed $F\subseteq P$ with $|F|\le t$, the number of faults in a committee has mean at most $(1/3-\varepsilon)k$. Exceeding the budget $\gamma k$ requires exceeding this mean by more than $\varepsilon k/2$. Applying \cref{lem:wor} with $y=\varepsilon k/2$ bounds this probability by $e^{-\varepsilon^2k/2}$.

We now bound the probability that the first property fails. If $\beta=1$, the first property is trivial, so assume $\beta<1$. Let $s=\lfloor\beta m\rfloor+1$, the least integer larger than $\beta m$. Then $\max\{1,\beta m\}\le s\le m$ and $m/s<1/\beta$, so $\binom ms\le(em/s)^s\le(e/\beta)^s$. The first property fails only if, for some fault set, some $s$ committees all exceed their fraction. There are at most $2^n$ fault sets and $(e/\beta)^s$ choices of these nodes; for each choice, independence bounds the probability that all $s$ committees are overloaded. The union bound is therefore
\[
2^n\Big(\frac e\beta\Big)^se^{-\varepsilon^2ks/2}.
\]
This is at most $1/4$ whenever $\varepsilon^2ks/2\ge n\ln2+\ln4+s\ln(e/\beta)$, and this follows from $k\ge K(m)$ because $s\ge\max\{1,\beta m\}$.

For the membership bound, observe that the number of memberships of a fixed process is binomial with mean $\mu=mk/n$. By \cref{lem:binom-upper} with $d=\Delta\ge\max\{2e\mu,3L\}$ (if $d$ exceeds the number of indicators, the event is impossible), it exceeds $\Delta$ with probability at most $2^{-3L}\le n^{-3}$. A union bound over the $n$ processes gives probability at most $1/4$. Hence some assignment has both properties.

To make the assignment explicit, every process enumerates all assignments of $k$-subsets to the ordered nodes of the level. It tests both properties, the first one for every $F\subseteq P$ with $|F|\le t$, and uses the first valid table in public lexicographic order. The search terminates by the existence argument and uses no communication. The table is public and is reused across epochs.

\subsection{Proof of \texorpdfstring{\cref{lem:uplinks}}{the forwarding-graph lemma}}\label{app:uplinks}

The proof follows the same pattern as for committees. If $D=k'$, we take the complete bipartite graph between $U$ and $V$. Then $|N(u)\cap Z|/D=|Z|/k'$ for every row $u$, so \eqref{eq:sampler} holds with no exceptional rows, and every column has degree $k$.

Otherwise, we give each row an independent uniform $D$-subset of $V$. Fix $Z\subseteq V$. By \cref{lem:wor}, with $\rho D$ in the role of $y$, a row violates the inequality in \eqref{eq:sampler} with probability at most $e^{-2D\rho^2}$. Let $s=\lfloor\rho k\rfloor+1>\rho k$. For a fixed set of $s$ rows, independence bounds the probability that they all violate the inequality. There are at most $2^{k\prime}$ choices of $Z$ and $2^k$ choices of the row set. Thus the probability that any $Z$ has $s$ violating rows is at most
\[
2^{k'}2^ke^{-2D\rho^2s}\ \le\ 2^{k+k'}e^{-2D\rho^3k}\ \le\ \frac18,
\]
where the last inequality is the assumption $2D\rho^3k\ge(k+k')\ln2+\ln8$.

A column's degree is a sum of $k$ independent indicators with mean $kD/k'$. By \cref{lem:binom-upper} with $d=J_{\mathrm{graph}}$ (if $d$ exceeds the number of indicators, the event is impossible), it exceeds $J_{\mathrm{graph}}$ with probability at most $2^{-3L}\le n^{-3}$. A union bound over at most $n$ columns gives at most $1/4$. Thus a graph with both properties exists, and it is fixed by exhaustive search.

\section{Parameter checks}\label{app:checks-packed}

We check that the parameters \eqref{eq:params-packed} satisfy the hypotheses used in the proof of \cref{thm:packed}.

\paragraph{Why these parameters.} The tree has at most $H$ levels, and $H$ is known before $w$ is fixed. The base committee size meets the committee bound, and later sizes grow geometrically, with an early cap at $n$. This controls the all-to-all voting cost.

\paragraph{Committees.} Assume $h\ge1$. Recall that $\kappa=\lceil K(m_1)\rceil$ is a base size, and that $k_\ell=\kappa b^{\ell-1}$ if $\kappa b^{\ell-1}\le n/b$ and $k_\ell=n$ otherwise. If $\beta m_\ell\ge1$, then also $\beta m_1\ge1$, and $K(m_\ell)\le b^{\ell-1}K(m_1)\le\kappa b^{\ell-1}$, because $m_\ell=m_1/b^{\ell-1}$; so $k_\ell\ge K(m_\ell)$ or $k_\ell=n$. If $\beta m_\ell<1$, then $\kappa b^{\ell-1}>n/b$: either $\beta m_1<1$ and $\kappa\ge K(m_1)>(2/\varepsilon^2)n\ln2>n$, or $\beta m_1\ge1$ and $\kappa b^{\ell-1}\ge(2/\varepsilon^2)\,n\ln2\cdot b^{\ell-1}/(\beta m_1)=(2/\varepsilon^2)\,n\ln2/(\beta m_\ell)>n$. So $k_\ell=n$. Either way \cref{lem:committees} applies. Since $m_h=1$ and $\beta<1$, the second case gives $k_h=n$, so the root committee is $P$, as the selection requires.

\paragraph{Small $n$ and $h=0$.} Since $w\ge A>10^6$, we have $h\ge1$ only if $10n>w$. Then $n>A/10>b$, so $\log_2b<L$, and $\log_2w<\log_2(10n)\le L+4$. If $h=0$, the root list is the single public list of all $N=w\ge10n$ arrays, there are no elections and no transfers, and every final word is dealt to $P$ by $V_{n,\sigma(n),1}$, a single-block sharing with threshold $\sigma(n)\ge t$; \cref{lem:moving-coins} then holds as in its proof.

\paragraph{Forwarding graphs and capacity.} We check four requirements in order: the graph exists, its incoming degree fits in $J$, a block has enough outputs to decode its packed inputs, and its corruption threshold leaves the slack required for path protection. Consider a transfer out of level $\ell$, so $k=k_\ell<n$, and let $k'=k_{\ell+1}$. Then $\kappa\le k\le n/b$ and $bk\le k'\le b^2k$: either $k'=bk$, or $k'=n$, and then $\kappa b^\ell>n/b$ gives $n/k<b^2$. 

In particular level~$1$ is not capped, so $\beta m_1\ge1$, and $m_1=N/(wb)<10n/w$ by the minimality of $h$. Hence
\[
k\ \ge\ \kappa\ \ge\ \frac{2}{\varepsilon^2}\cdot\frac{n\ln2}{\beta m_1}\ >\ \frac{4\ln2}{\varepsilon^2}\,Hw\ \ge\ \frac{4\ln2}{\varepsilon^2}\,AH^3L\ >\ 2C_\varepsilon L+2,
\]
using $A=10^6(c+2)\varepsilon^{-2}$ and $C_\varepsilon\le2000\varepsilon^{-3}+1$. So $C_\varepsilon Lk'/k<k'/2$, and $D_\ell<k'$: the forwarding graph is not complete. Moreover $2D_\ell\rho^3k\ge4Lk'\ge(k+k')\ln2+\ln8$, so \cref{lem:uplinks} applies. 

The graph has mean column degree $kD_\ell/k'$ lies between $C_\varepsilon L$ and $C_\varepsilon L+1/b$, so the graph-specific bound is $J_{\mathrm{graph}}=\lceil2ekD_\ell/k'\rceil\le2eC_\varepsilon L+2e/b+1\le6C_\varepsilon L\le J$. Thus a holder at level $\ell\ge2$ has at most $J$ registers of a word. Since $D_\ell\ge bC_\varepsilon L$ and $J\le7C_\varepsilon L$, we get
\[
p_\ell\le J\le\tfrac7bD_\ell\le0.7\,\varepsilon D_\ell<\tfrac{3\varepsilon}4D_\ell\le D_\ell-3\sigma(D_\ell),
\]
where the last step uses $1-3\theta=3\varepsilon/4$; this is the decoding condition of \cref{lem:moving-coins}. Blocks with $p=1$ satisfy it because $3\sigma(D)=3\lfloor\theta D\rfloor<D$. 

For \cref{lem:path}, $D_\ell\ge bC_\varepsilon L>20/\varepsilon$ gives $\gamma+2\rho+D_\ell^{-1}<\theta$, and $k_1=\kappa\ge8/\varepsilon^2$ gives $(\theta-\gamma-\rho)k_1=0.15\,\varepsilon k_1\ge1$, hence $(\gamma+\rho)k_1\le\theta k_1-1<\sigma(k_1)$. If the level-$1$ committees are $P$, there are no transfers and $\sigma(n)\ge t$. So \cref{lem:moving-coins} applies.

\paragraph{Root inequality.} We have $h\le H$, and by the bound after \cref{prop:root}, $F_0\ge\frac23+\varepsilon-t/N\ge\frac23+\varepsilon-\frac1{30}$ on every execution, since $N\ge10n$ and $t<n/3$. With $h(\beta+\tau+\eta)\le\frac1{20}+\frac1{100}+\frac1{20}$ this gives
\[
F_0-h(\beta+\tau+\eta)\ \ge\ \tfrac{157}{300}+\varepsilon\ >\ 0.52 .
\]

\paragraph{Failure probability.} We make the failure probability of each election at most $n^{-(c+2)}/2$. There are at most $n$ elections, leaving room for their union bound and the final-agreement failure probability within $n^{-c}$. If $h=0$, there are no elections, $\delta=0$, and $p_n=2^{-w/2}\le n^{-(c+2)}$. Let $h\ge1$. Then $N<10nb$ by the minimality of $h$, so $\iota\le\log_2N+2\le2L+6$, and $\log_2u=\log_2b+\log_2w+\log_2(\iota+\log_2b)\le L+(L+4)+(L+4)=3L+8$. Since $\tau r=bw/(100H)\ge bAHL/100\ge(c+5)L+10$, we get $u\,2^{-\tau r}\le2^{-(c+2)L-2}\le n^{-(c+2)}/4$. Also $\eta^2w/2=w/(800H^2)\ge AL/800\ge(c+2)L+4+\ln b$, so $b\,e^{-\eta^2w/2}\le e^{-4}n^{-(c+2)}$, using $e^{-L}\le1/n$. Hence $\xi\le n^{-(c+2)}/2$. 

The tree has $\sum_\ell m_\ell\le2m_1<20n/w\le n$ internal nodes, so $\delta\le n^{-(c+1)}/2$ by \eqref{eq:delta}. Finally $w/2\ge(c+2)L$ gives $2^{-w/2}\le n^{-(c+2)}$, so $p_n\le n^{-c}$.

\paragraph{Cost.} \Cref{app:costs} shows that one epoch costs $O_{\varepsilon,c}(nL^8)$ qubits, $O_{\varepsilon,c}(n^2L^8)$ classical bits and $O_{\varepsilon,c}(L^4)$ rounds, deterministically. The rounds number $1+h(4bw+3)+4(w+1)=O(Hbw)$, with $w=\Theta_{\varepsilon,c}(H^2L)$ and $H=O(L)$.

\section{Qubits of the Ben-Or--Hassidim coin}\label{app:bh}

Ben-Or and Hassidim~\cite{BenOrHassidim05} do not bound the communication of their Byzantine protocol. Its coin follows the oblivious coin of Feldman and Micali~\cite{FeldmanMicali97}: every process $k$ shares a secret for every other process $i$, and the secrets are later opened and summed modulo $n$ \cite[Section~5]{BenOrHassidim05}. Each secret is the uniform superposition $n^{-1/2}\sum_{a=0}^{n-1}\ket a$, and it is shared by a graded variant of the verifiable quantum secret sharing of Cr\'epeau, Gottesman and Smith~\cite{CrepeauGottesmanSmith02}, which shares the secret using a two-layered variant of polynomial codes; for $t<n/3$ the polynomials have degree $t$ and the phase check is skipped \cite[Section~5]{BenOrHassidim05}. We count the qubits of one coin flip.

\begin{proposition}\label[proposition]{prop:bh}
In an execution without corruptions, one coin flip of the Byzantine protocol of~\cite{BenOrHassidim05} sends at least $n^2(n-1)^2\log_2n=\Omega(n^4\log n)$ qubits.
\end{proposition}

\begin{proof}
One coin flip has $n(n-1)$ sharings, one for each ordered pair of distinct processes. In the two-layered sharing of~\cite[Protocol~1]{CrepeauGottesmanSmith02}, the dealer encodes the shared system with a polynomial code of length $n$ and sends the $i$th component to process $i$. Each process then encodes the component it received again, with the same code, into $n$ qudits, and sends the $j$th of them to process $j$. For the shared system alone, this second layer sends at least $n(n-1)$ qudits; the systems used in the verification only add to this. A qudit that carries a superposition over $n$ values has dimension at least $n$, so it takes at least $\log_2n$ qubits. The three factors are therefore
\[
\underbrace{n(n-1)}_{\text{sharings}}
\cdot\underbrace{n(n-1)}_{\text{second-layer transmissions per sharing}}
\cdot\underbrace{\log_2 n}_{\text{qubits per transmitted system}},
\]
which gives the bound. Counting this layer alone suffices; all other transmissions only increase the cost.
\end{proof}

The bound uses only the two-layered structure of the sharing, not the parameters of its verification. Cr\'epeau, Gottesman and Smith also describe an optimized version in which each player sends $O(n+\log(1/\epsilon_{\mathrm{snd}}))$ qudits per sharing, where $\epsilon_{\mathrm{snd}}$ is the soundness error~\cite{CrepeauGottesmanSmith02}. If the verification is implemented in this way, with $\epsilon_{\mathrm{snd}}=n^{-O(1)}$ as a union bound over the $n(n-1)$ sharings requires, then one coin flip sends $\Theta(n^4\log n)$ qubits.

\newcommand{\etalchar}[1]{$^{#1}$}

\end{document}